\documentclass{aa}

\usepackage{graphicx}
\usepackage{makecell}
\usepackage{txfonts}
\usepackage{booktabs}
\usepackage{natbib}
\usepackage{svg}
\usepackage{placeins}
\usepackage{tikz}
\usepackage{hyperref}
\usepackage[utf8]{inputenc}
\usepackage[T1]{fontenc}
\usepackage{orcidlink}
\usepackage{ulem}
\usepackage{bm}
\usepackage{xcolor}

\newcommand{\gttt}{G332}
\newcommand{\gte}{G28}
\newcommand{\gft}{G14}
\newcommand{\unit}[1]{\ensuremath{\,\mathrm{#1}}}
\newcommand{\msun}{\mathrm{M}_\odot}
\newcommand{\lsun}{\mathrm{L}_\odot}
\newcommand{\pillaiOverviewPaper}{Pillai et al., in prep.}

\begin{document}

   \title{Before fragmentation: from uniform clouds to stellar clusters}

   \author{A.~Giannetti\corrauth{andrea.giannetti@inaf.it}
          \inst{1,2\orcidlink{0000-0003-3869-6501}}
          \and
          T.~Pillai\inst{3}
          \and
          J.~Kauffmann\inst{3\orcidlink{0000-0002-5094-6393}}
          \and
          A.~Palau\inst{2}
          \and
          E.~V\'azquez-Semadeni\inst{2}
          \and
          K.~Morii\inst{4,5}
          \and
          P.~Sanhueza\inst{4\orcidlink{0000-0002-7125-7685}}
}

\institute{INAF - Istituto di Radioastronomia di Bologna, Via Gobetti 101, 40129 Bologna, Italy\\
    \email{andrea.giannetti@inaf.it}
    \and
    Instituto de Radioastronomía y Astrofísica (IRyA-UNAM), 3-72 (Xangari), 8701, Morelia, Mexico
    \and
    Haystack Observatory, Massachusetts Institute of Technology, 99 Millstone Road, Westford, MA 01886, USA
    \and
    Department of Astronomy, School of Science, The University of Tokyo, 7-3-1 Hongo, Bunkyo, Tokyo 113-0033, Japan
    \and
    National Astronomical Observatory of Japan, National Institutes of Natural Sciences, 2-21-1 Osawa, Mitaka, Tokyo 181-8588, Japan
}
  \abstract
   {
   Our systematic study of the early stages of fragmentation in massive clumps has revealed a striking anomaly: $\sim15\%$ of the Cold Cores with ALMA (CoCoA) survey sample exhibit no fragmentation at all. These sources could represent the initial conditions before core formation.
   }
   {
   We set out to investigate whether these unfragmented clumps are indeed at the onset of core and cluster formation, and whether gravity or turbulence dominates their early fragmentation.
   }
   {
   To understand the nature of these extreme sources, we study the transition from uniform gas to active cluster formation, and compare contrasting predictions of turbulence- and gravity-dominated models of clump fragmentation. We used ALMA observations of three massive clumps spanning this phase (from two minimally fragmented clumps, to one significantly fragmented source).
   Our analysis includes direct, gas-phase volume density estimates from CH$_3$OH$(2_K - 1_K)$ line ratios (independent of dust properties, temperature, and geometry), unfiltered kinematics of the dense gas (CH$_3$OH and $\mathrm{N_2H^+} J=1-0$), and a comparison between the mass build-up- and chemical timescales derived from methanol abundances.
   }
   {
   The fragmented source exhibits higher average volume densities and a core-to-envelope density contrast $5-6$ times larger than the minimally fragmented sources. All sources display pc-scale velocity differences comparable to their linewidths, and both are similar to the gravitational velocity. This suggests that the linewidth is dominated by the large-scale systematic motions, and that the latter are driven by gravity. We also estimate high CH$_3$OH abundances ($\sim$10$^{-9}$) across the clumps. These abundances and the inferred mass build-up timescales align with astrochemical models where complex molecules form at intermediate densities before the appearance of cores.
   }
   {
    All independent physical and chemical diagnostics in our analysis are consistent with a scenario where the minimally fragmented clumps represent an early evolutionary phase, potentially marking the onset of core fragmentation driven by gravitational collapse.
    If turbulence were the primary driver of fragmentation, sources with comparable Mach numbers ($\mathcal{M} \approx 6-8$ in our sample) should exhibit similar structural properties. Instead, we observe significant variations in density contrast and fragmentation. These findings challenge purely turbulence-supported models and lend support to hierarchical, gravity-driven models of clump fragmentation, where large-scale ordered motions account for a significant fraction of the observed linewidths and drive the evolution of the clump toward increasingly dense and fragmented states.
}

   \maketitle
   \nolinenumbers

\section{Introduction}

While high-mass star formation (SF) theories qualitatively describe the transformation of molecular gas into stars, several open questions persist. These include the mechanisms driving and regulating the process, the observed low star formation efficiency, and the interplay between gravitational collapse and feedback.
Recent ALMA observations strongly favour clump-fed models of SF, such as competitive accretion \citep{bonnell_star_2006}, the global hierarchical collapse model \citep[GHC, ][]{vazquez-semadeni_global_2019}, and the inertial-inflow scenario \citep{padoan_origin_2020}, over core-fed alternatives \citep[e.g.][]{sanhueza_massive_2017, contreras_infall_2018, sanhueza_alma_2019, morii_alma_2021, li_alma_2023, morii_alma_2024, wallace_almagal_2026}.
However, further scrutiny is required to determine whether turbulence or gravity drives early clump fragmentation.

The Cold Cores with ALMA (CoCoA; \pillaiOverviewPaper) survey systematically targets quiescent, massive ATLASGAL \citep[APEX\footnote{Atacama Pathfinder EXperiment, \citet{gusten_atacama_2006}} Telescope Large Area Survey of the Galaxy][]{schuller_atlasgal_2009} clumps (typically with $R\sim 0.5\unit{pc}$ and $M\sim 100-1000\unit{M_\odot}$; \citealt{urquhart_atlasgal_2018}) in the inner Galactic plane within $3-5 \unit{kpc}$. It avoids selection biases towards the brightest, earliest sources, thereby providing a complete view of early cluster formation.
CoCoA observed 238 clumps at $230\unit{GHz}$ using the ALMA main array (MA) and Atacama Compact Array (ACA). The combined dataset achieves $1.5\arcsec$ ($\approx0.04\unit{pc}$) resolution and a maximum recoverable scale (MRS) of $\approx 20\arcsec$.
Around $15\%$ of the 238 CoCoA sample exhibits no compact continuum emission at the level of cores. Throughout this work, unfragmented or lacking fragmentation refer to the non-detection of compact $1.3\unit{mm}$ continuum cores above our $5\sigma$ sensitivity limit in the combined MA+ACA images ($\lesssim0.2\unit{M_{\odot}/beam}$, derived assuming a $15\unit{K}$ temperature and using the most conservative beam size of $1.6\unit{\arcsec}$), indicating that the mass reservoir is smoothly distributed across the pc-scale intraclump medium rather than divided into discrete cores.
This raises a compelling question: where has the vast reservoir of material, amounting to hundreds of $\msun$ visible at pc-scale, vanished to? The lack of compact continuum emission suggests that the clumps are not fragmented into centrally-concentrated cores. Indeed, \citet{sanhueza_alma_2019} propose that the large fraction of unrecovered flux in ASHES sources resides in a more uniform, intraclump medium.
A similar conclusion about the increase in the mm-continuum structural complexity is reached by \citet{wallace_almagal_2026}, which show how higher temperatures, surface densities, and $L/M$ ratios are preferentially connected to sources with highly-nested dendrogram hierarchies.

Fragmentation encodes information on the processes shaping SF.
The $\sim 15\%$ of completely unfragmented clumps represent an extreme case where the emission is so extended and uniform that nothing is detected above $5\sigma$ in the CoCoA data.
This lack of substructure may represent the initial stages of gravitational contraction or may be generated by intrinsically different initial properties of the sources.
Specifically, classical turbulent support frameworks assume that the cloud and its substructures are maintained in an approximate virial equilibrium between gravity and turbulence \citep{mac_low_control_2004, bergin_cold_2007, mckee_theory_2007, hennebelle_turbulent_2012, vazquez-semadeni_turbulent_2026}. In this scenario, while turbulence dynamically generates local density fluctuations via shocks \citep{vazquez-semadeni_compressible_2000, padoan_turbulent_2001}, its primary role is to provide support to counteract rapid gravitational contraction \citep{zuckerman_models_1974, mac_low_control_2004}. This framework predicts a roughly time-independent star formation rate regulated by the initial turbulent parameters of the gas, rather than a steady structural and density evolution over time \citep{federrath_star_2012, vazquez-semadeni_turbulent_2026}.
In contrast, gravitational collapse with moderate turbulent density fluctuations at clump scales allows for an initial phase with minimal substructure \citep{vazquez-semadeni_global_2019, vazquez-semadeni_turbulent_2026} because ordered, large-scale flows do not produce strong shocks \citep[e.g.][]{gomez_filaments_2014, naranjo-romero_gravity-driven_2022} and the density contrast is initially low.
The collapse occurs from the outside-in, so that the substructures of a large-scale structure undergoing global contraction become locally unstable, and begin to grow themselves later than the parent structure \citep{vazquez-semadeni_global_2019}.
The collapsing regions evolve toward greater density and fragmentation as they accumulate mass and collapse progresses.
Therefore, to understand whether their lack of substructure is a permanent feature caused by their specific properties or a transient initial phase, it is important to study the transition between uniform gas and a fully-developed core population.

While CoCoA continuum observations show this varied clump substructure, they do not allow us to thoroughly characterise the volume density distribution and investigate its potential evolution.
Methanol offers a powerful alternative for studying it.
The integrated-intensity ratios of its $(2_K - 1_K)$ band scale monotonically with density while remaining largely insensitive to temperature.
These ratios constrain the line-of-sight (LOS)-averaged density to within a factor of two to three for $n(\mathrm{H_2}) \sim 5 \times 10^4 - 3 \times 10^7$\,cm$^{-3}$, which is typical of intraclump and core gas \citep{giannetti_ch3oh_2025}.

Thus, we exploit our newly-calibrated density probe to analyse the three CoCoA sources in the ALMA archive with available methanol Band 3 data.
They cover the critical evolutionary step from diffuse clumps showing only the first, minimal signs of fragmentation (serving as the closest observational link to the unfragmented extremes) to a system already breaking into numerous compact continuum cores.
By comparing the molecular gas density distribution and kinematics of the sources with the theoretical predictions for gravitational and turbulent fragmentation, we aim to establish whether the extreme, unfragmented sources represent the initial conditions of high-mass cluster formation.

\section{Observations and methods}\label{sec:obs_methods}

    \begin{table}[tb]
        \caption{\label{tab:lines_list}Rest frequencies of the lines considered in this work.}
        \small
        $$
        \begin{array}{cc}
            \hline
            \hline
            \mathrm{Transition}            & \mathrm{Rest \, Frequency}          \\
                                           & [\mathrm{GHz}]                      \\
            \hline
            \mathrm{E-CH_3OH(2_{-1} - 1_{-1})}   & 96.739358                    \\ 
            \mathrm{A-CH_3OH(2_{0}^+ - 1_{0}^+)} & 96.741420                    \\
            \mathrm{E-CH_3OH(2_{0} - 1_{0})}     & 96.744545                    \\
            \mathrm{E-CH_3OH(2_{1} - 1_{1})}     & 96.755501                    \\
            \mathrm{N_2H^+(1 - 0)}               & 93.173398                    \\ 
            \hline
        \end{array}
        $$
        \tablefoot{Rest frequencies are taken from the JPL database \citep{pickett_submillimeter_1998} for CH$_3$OH and from the Cologne Database for Molecular Spectroscopy \citep{mueller_cologne_2001, mueller_cologne_2005} for N$_2$H$^+$. Primary laboratory measurements are based on \citet{xu_microwave_1997} for CH$_3$OH and \citet{caselli_radio_1995} for N$_2$H$^+$.}
    \end{table}

    \begin{table*}[th]
        \centering
        \small
        \caption{Properties of the CoCoA sources from the ATLASGAL catalogue \citep{urquhart_atlasgal_2022}.}
        \label{tab:cocoa_sources}
        \begin{tabular}{lrrrrrrrrr}
            \toprule
            Source & Distance & Mass & $T_{dust}$ & Luminosity & $L/M$ ratio & $R_{eff}$ & $F_\mathrm{peak}^{870\mu m}$ & V$_{lsr}$ & $N(\mathrm{H_2})$ \\
                   & (kpc)     & ($\msun$) & (K) & ($\lsun$) & ($\msun \, \lsun^{-1}$) & ($\mathrm{pc}$) & ($\mathrm{Jy \, beam^{-1}}$) & ($\mathrm{km \, s^{-1}}$) & ($10^{22}\times\mathrm{cm^{-2}}$) \\
            \midrule
            AGAL332.969-00.029 & 4.0 & 1330 & 12.2 & 90 & 0.07 & 0.42 & 0.53 & -67.6 & 2.1 \\
            AGAL028.273-00.167 & 4.5 & 2250$^a$ & 11.8 & 790 & 0.32 & 0.40$^a$ & 1.02 & 79.6 & 3.7 \\
            AGAL014.492-00.139 & 3.1 & 2050 & 16.6 & 3420 & 1.67 & 0.26 & 2.70 & 40.5 & 4.6 \\
            \bottomrule
        \end{tabular}
        \tablefoot{
            The sources are ordered according to ascending $L/M$ ratio and fragmentation level, as a proxy of their evolution.
            The columns indicate the distance, mass, dust temperature ($T_{dust}$), bolometric luminosity, luminosity-to-mass ratio ($L/M$), effective radius ($R_{eff}$), peak $870\unit{\mu m}$ flux ($F_\mathrm{peak}^{870\mu m}$), systemic radial velocity (V$_{lsr}$), and peak H$_2$ column density ($N(\mathrm{H_2})$).
            $^a$ These values have been rescaled to a smaller aperture, to exclude a close-by active star-forming region in the ATLASGAL total flux that is not present in the ALMA observations.
        }
    \end{table*}

    \begin{table*}[th]
        \centering
        \small
        \caption{Properties of the final, feathered ALMA+TP cubes for $\mathrm{N_2H^+}(1-0)$ and $\mathrm{CH_3OH}(2_K - 1_K)$.}
        \label{tab:image_props}
        \begin{tabular}{l c c c r r r r}
            \toprule
            Source & Angular beam & Linear beam & Image px & Px size & Channels & $\Delta\varv$ & $\sigma_{spec}$ \\
            & ($\arcsec\times\arcsec$) & ($\mathrm{AU} \times \mathrm{AU}$) & & ($\arcsec$) & & ($\mathrm{km \, s^{-1}}$) & $\left(\frac{mJy}{beam}\right)$ \\
            \midrule
            AGAL332.969-00.029 & $1.8\times 1.6$ & $7200\times 6400$ & $512 \times 512$ & 0.2 & 500 & 0.3 & 3.4 \\
            AGAL028.273-00.167 & $2.8 \times 1.5$ & $12600 \times 6750$ & $512 \times 512$ & 0.2 & 500 & 0.3 & 3.0 \\
            AGAL014.492-00.139 & $2.3\times 1.5$ & $7130\times 4650$ & $512 \times 512$ & 0.2 & 500 & 0.3 & 3.2 \\
            \bottomrule
        \end{tabular}
        \tablefoot{``Px size'' indicates the angular size of the square pixels in the final image, ``$\Delta\varv$'' is the channel width in the spectra, and the RMS noise per pixel of the final spectra is reported as $\sigma_{spec}$, being essentially identical in both datacubes.}
    \end{table*}

A search of the ALMA archive for the CoCoA sources in the 3~mm window at the frequency of CH$_3$OH$(2_K - 1_K)$ band yielded three sources: AGAL332.969-00.029 (\gttt), AGAL028.273-00.167 (\gte), and AGAL014.492-00.139 (\gft).
Table~\ref{tab:cocoa_sources} summarises the key properties of these clumps from ATLASGAL.

\subsection{The sample}
In the CoCoA morphological classification (\pillaiOverviewPaper), which categorises clump
fragmentation based on dendrogram-derived structural properties of the 1.3\,mm continuum emission, \gte\ and \gttt\ fall in the Diffuse category, fields dominated by faint, extended emission with few
prominent compact cores, while \gft\ is classified as Structured, exhibiting multiple bright cores embedded in a hierarchical substructure. All three sources have been observed as part of the GLASHES survey \citep{morii_global_2025}, in addition to ASHES.
\gte\ and \gft\ have been the subject of dedicated interferometric studies, which provide crucial context for their evolutionary state.
The main clump in \gte\ (also known as IRDC G028.23-00.19) has been presumed to be in a prestellar state, based on its low dust temperature, narrow molecular linewidths, and a lack of embedded infrared sources up to $70\unit{\mu m}$ \citep{sanhueza_distinct_2013}, and it is one of the few to retain this classification after high-resolution interferometric observations \citep{sanhueza_massive_2017}. In fact, both SiO and CO fail to reveal even molecular outflows from the few low-mass cores identified in the mm continuum, which are the first signposts of SF \citep{urquhart_atlasgal_2022}.
Based on the velocity structure of optically thin tracers, \citet{sanhueza_distinct_2013} suggest that the main clump might be still accreting mass from its environment.
\gft\ has also been classified as a $70\unit{\mu m}$-dark source, but high-sensitivity ALMA observations have revealed localised signs of SF. \citet{li_alma_2020} identified multiple bipolar outflows associated with dense cores, indicating the presence of embedded protostars. Furthermore, \citet{redaelli_core_2022} characterised the prestellar core population using o-H$_2$D$^+$, identifying 22 cold cores with masses $<30\unit{M_\odot}$. Their kinematic analysis of N$_2$H$^+$ revealed a filamentary structure with velocity gradients consistent with gas accreting onto a protostellar core at a rate of $\sim 2\times 10^{-4}\unit{M_\odot \, yr^{-1}}$, supporting a clump-fed growth scenario.
In contrast, \gttt\ lacks comparable, dedicated, high-resolution studies, though it is identified as a massive clump in the early stages of evolution within the ATLASGAL and CoCoA surveys, and no CO outflows have been detected by \citet{li_alma_2020}.

\subsection{ALMA archival data}
The GLASHES Band 3 data include the MA, ACA, and Total Power (TP) antennas (project 2018.1.00299.S, PI: Contreras).
The data cover the N$_2$H$^+(1-0)$ line, in addition to the CH$_3$OH$(2_K - 1_K)$ band; the rest frequencies of these transitions are listed in Table~\ref{tab:lines_list}.
We used standard ALMA archive scripts for data reduction and reviewed calibration reports to ensure good data quality.
We combined the visibilities of the calibrated ACA and MA measurement sets for joint deconvolution.
We performed imaging using the \texttt{tclean} task in CASA \citep{the_casa_team_casa_2022}, with automasking down to a threshold of $3.5 \unit{mJy\,beam^{-1}}$ to recover extended emission.
We then combined the single-dish data via the feathering technique.
Combining all arrays removes the limitations imposed by the MRS characteristic of purely interferometric data, thus avoiding the filtering issues that affect the CoCoA continuum.
To assess the fidelity of this process, we compared the combined-cubes spectra directly against the TP-only spectra (see Appendix~\ref{app:feathering_verification}), confirming that the flux recovery is complete and that kinematics are indeed unfiltered.
The properties of the final cubes are listed in Table~\ref{tab:image_props}.

\begin{figure*}[th]
    \centering
    \includegraphics[width=0.9\textwidth]{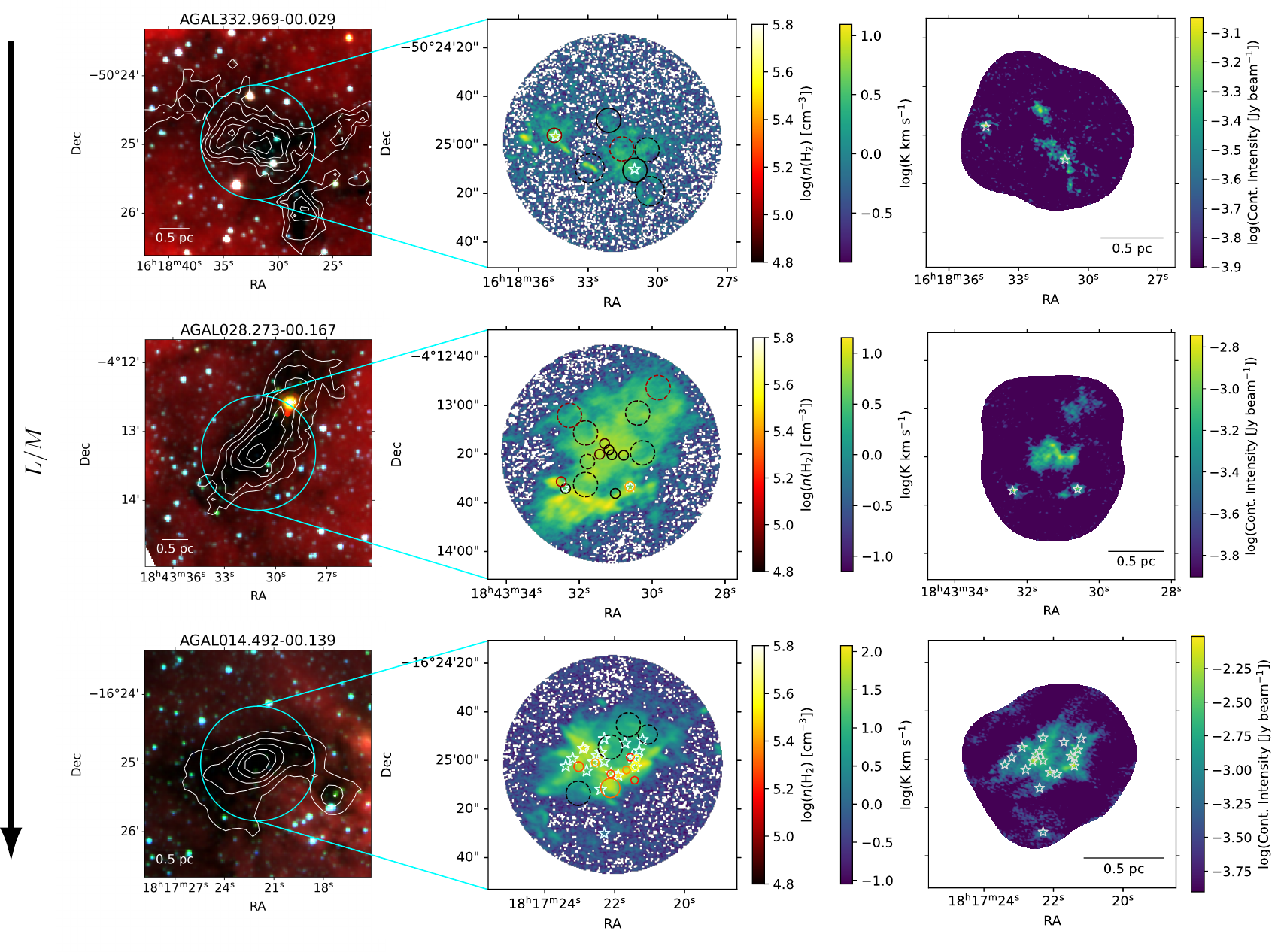}
    \caption{Comparison of pc-scale and high-resolution data. Left: ATLASGAL contours (seven levels from $0.1\unit{Jy\,beam^{-1}}$ to maximum; see Table~\ref{tab:cocoa_sources}) on top of the GLIMPSE RGB image. Middle: CH$_3$OH $(2_K-1_K)$ moment-zero with colour-coded densities for each region (only the most reliable component; solid circles indicate the cores, and dashed ones the envelope regions; cf. Sect.~\ref{sec:line_fitting} and Tables~\ref{tab:G332.96_densities}, \ref{tab:G28_densities}, and \ref{tab:G14_densities}). Right: ASHES continuum. For comparison, actively star-forming cores from the catalogue of \citet{morii_alma_2024} are indicated by white stars.}\label{fig:main}
\end{figure*}

\subsection{Line fitting and region selection}\label{sec:line_fitting}
We fitted CH$_3$OH and N$_2$H$^+$ on a pixel-by-pixel basis to map the velocity field of the clumps.
Our fitting pipeline is capable of fitting up to three velocity components along the line of sight, a number that was selected via the direct inspection of the data, and validated a posteriori.
We adopted a progressive, dynamically-masked fitting strategy using the Bayesian Information Criterion (BIC).
The routine iteratively tests models with an increasing number of components.
After an $N$-component fit, the addition of an $N+1$ component is limited to a progressively smaller spatial area where the BIC of the current residuals strictly exceeds a theoretically derived $3\sigma$ noise threshold, which indicates that statistically significant residuals remain in the spectrum.
When fit results with a different number of components are available for a pixel, the more complex one is only accepted if it reduces the BIC compared to the simpler model, balancing the goodness-of-fit against the penalty for additional free parameters.
Finally, to ensure the physical reliability of the selected components, we applied validation checks to the resulting fit parameters.
A velocity component is only retained if its peak intensity exceeds the $3\sigma$ threshold, and the relative uncertainty of its fitted parameters are found to be below $50\%$.
Any component that does not satisfy these criteria, or that have unconstrained parameters, is automatically excluded from the subsequent kinematic and density analyses.

We also extracted the spectra from representative regions to determine gas densities (see Tables \ref{tab:G332.96_densities}--\ref{tab:G14_densities}).
To characterise the density contrast within the clumps, we classified extracted regions into three types:
\begin{itemize}
    \item core\_cont: Positions centred on compact $1.3\unit{mm}$ continuum peaks;
    \item core\_ch3oh: Significant methanol emission peaks found offset from the continuum (excluding known outflow knots identified in H$_2$CO; \citealt{izumi_alma_2024});
    \item envelope: Areas representing the diffuse, inter-core medium, selected where emission is spatially uniform and lacks compact structures.
\end{itemize}
Aperture sizes were chosen to maximise the signal-to-noise ratio (SNR) of the weak CH$_3$OH $(2_0-1_0)$ line.
While sizes vary slightly to accommodate local emission morphology, they remain generally consistent across the sample.
Finally, we extracted the spectrum over the entire clump, to compute their average density (see Table~\ref{tab:results_summary}).

For the fitting itself, we simultaneously modelled methanol's four $(2_K-1_K)$ lines with Gaussians using PySpecKit \citep{ginsburg_pyspeckit_2011, ginsburg_pyspeckit_2022}, fixing their radial velocity separation and full width at half maximum (FWHM) to be the same. We set a minimum peak brightness temperature of $1\sigma$ to easily identify non-detections in the weakest lines (see Sect.~\ref{sec:density_method}).
We only fitted the isolated hyperfine component of the N$_2$H$^+(1-0)$ transition, to reduce potential optical depth issues. For both species, we imposed a minimum FWHM of $0.5 \unit{km\,s^{-1}}$ (approximately two channels).
To improve decomposition in cases of multiple components and low SNR, we applied a stricter maximum linewidth of $3\unit{km \, s^{-1}}$ to the isolated N$_2$H$^+$ component fits. Visual inspection confirmed significant improvements, indicating no larger linewidth was required.

Unlike N$_2$H$^+$, which is a simple linear molecule whose population strongly concentrates in the lowest rotational states for the typical properties of quiescent clumps (often leading to high optical depths in the main hyperfine components), CH$_3$OH is an asymmetric top species. Its population is distributed across a large number of $K$-ladders and $A/E$ symmetry states. Because of this dilution, the fractional population in the upper states of the $96.7\unit{GHz}$ ($J=2_K-1_K$) band remains relatively small. RADEX tests confirm that under the typical physical conditions of our clumps ($T_{\mathrm{kin}} \approx 12-15\unit{K}$, $n(\mathrm{H_2}) \sim 5-10 \times 10^4 \unit{cm^{-3}}$, $\mathrm{FWHM} \sim 1.5\unit{km\,s^{-1}}$, and $N(\mathrm{CH_3OH})\lesssim 10^{14} \unit{cm^{-3}}$, these transitions remain optically thin ($\tau < 0.25$ for the most optically thick E-CH$_3$OH line). This is empirically supported by the distinct relative intensities of the $K$-components in our spectra; if the lines were optically thick, their intensity ratios would thermalise toward unity, which would prevent the density inference.

To address the complexity arising from multiple velocity components along the LOS, we used a Bayesian Gaussian Mixture clustering model to process the fitted radial velocities ($V_{\mathrm{LSR}}$) in Position-Position-Velocity (PPV) space. The Bayesian Gaussian Mixture model allowed the probabilistic identification and segregation of distinct kinematic structures. The number of clusters was optimised to isolate the main, physically contiguous velocity component, which corresponds to the main contributor to the targeted clump. This process was essential for the kinematic analysis of the velocity gradients in the clumps (Sect.~\ref{sec:kinematics}).

\subsection{Number density inference}\label{sec:density_method}
We determined the gas density using the PyMC package \citep{abril-pla_pymc_2023} to model the posterior distribution of the integrated line intensity ratios (both averaged over the entire clumps and in each of the apertures considered, see Sect.~\ref{sec:line_fitting} and Fig.~\ref{fig:main}) of the three E-methanol lines among the four that were fitted. This approach provides a robust evaluation of the uncertainty and the proper treatment of non-detections.

Our model uses a censored likelihood to handle spectra where a line is undetected (defined as having a peak $\mathrm{SNR} < 3$).
This incorporates the information that the true line intensity is below the $3\sigma$ detection limit without biasing the result with a low SNR measurement.
For detections, a standard Gaussian likelihood was used, centred on the observed integrated intensity.

The model samples the posterior of the intensity ratios using a Markov Chain Monte Carlo sampler. These posterior samples of the ratios were then converted into samples of gas density through the analytical approximations given in \citet{giannetti_ch3oh_2025}. The final probability density function (PDF) for the gas density was generated for each component using a Kernel density estimate. The best-fit density was taken as the peak of this PDF, and the uncertainty was defined by the 67\% highest-probability density (HPD) interval.

\subsection{Filamentary infall rate calculation}\label{sec:infall_method}
To estimate the filamentary infall rate ($\dot{M}$) onto the clump, we focus on \gte, which exhibits the clearest velocity field and bright methanol emission, enabling a robust size and kinematic evaluation (cf. Fig.~\ref{fig:main}).
A tentative estimate is also provided for \gttt, whereas we refrain from considering \gft\ because of the presence of two close-by filamentary structures that complicate this analysis.
We model the structures exhibiting a linear velocity gradient (Sect.~\ref{sec:kinematics}, Fig.~\ref{fig:velocity_fields}) as an inclined elliptical sheet of molecular material infalling towards its projected centre. We stress that this is a rough approximation, and not necessarily the true geometry of the structure. We assume the elliptical sheet is inclined by $\gamma = 30^\circ$ along the major axis and $\theta = 45^\circ$ along the minor axis. These angles are consistent with the configuration required to measure radial velocity gradients while avoiding unphysically extreme aspect ratios. A schematic representation of this model is shown in Fig.~\ref{fig:geometry}.

We estimate the total mass ($M$) in the elliptical sheet using $M=\rho V$. The deprojected, physical volume ($V$) of the inclined sheet is:
\begin{equation}
    V=\pi \left( \frac{L_p}{2 \cos \gamma} \right) \left( \frac{W_p}{2 \cos \theta} \right) H.
\end{equation}
Here, $L_p$ and $W_p$ are the projected major and minor axis diameters (hence the factor of 2 to obtain radii), and $H$ is the intrinsic physical thickness of the sheet.

We approximate the infall timescale, $\tau_{inf}$, as the dynamical time for material at the outer edge to reach the centre. By de-projecting both the physical radius ($R = W_p / 2 \cos \theta$) and the infall velocity ($\varv_{inf} \approx ||\nabla\varv_{rad}|| (W_p/2) \sin \theta$), we obtain:
\begin{equation}
    \tau_{inf} = \frac{R}{\varv_{inf}} = \frac{\tan \theta}{||\nabla\varv_{rad}||}.\label{eq:t_inf}
\end{equation}
A detailed derivation of the quantities discussed here is provided in Appendix~\ref{app:geometry}. We then calculate the final order-of-magnitude estimate for the infall rate ($\dot{M}$) from the environment onto the filamentary clump using $\dot{M} = M / \tau_{inf}$.
The main sources of uncertainty in this estimate are the volume density derivation from CH$_3$OH (a factor of $2-3$; see Sect.~\ref{sec:density_method}) and the assumed inclination angles. However, the final filamentary infall rate is robust against the exact orientation.
   Because the intrinsic thickness is derived from the observable LOS path length ($H = H_{\mathrm{LOS}} \cos \gamma \cos \theta$), the angular dependencies perfectly cancel out in the volume calculation ($V \propto L_p W_p H_{\mathrm{LOS}}$), making the clump mass independent of inclination.
   As a result, the geometric uncertainty of the accretion rate depends only on the timescale ($\tau_{inf}$). Because the timescale is the ratio of the deprojected physical radius and the deprojected velocity, it scales as $\tan \theta$ (Eq.~\ref{eq:t_inf}). The final accretion rate therefore scales simply as $\cot \theta$, removing the dependence on the major axis inclination $\gamma$. Varying $\theta$ across a broad, realistic range of $30^\circ$ to $60^\circ$ changes $\dot{M}$ by less than a factor of $2$ relative to the fiducial value at $\theta = 45^\circ$ (a total range of a factor of $\sim 3$ across the interval). Therefore, the combined uncertainties confirm that our calculation reliably captures the order of magnitude of the environmental mass supply.

\section{Results}\label{sec:results}
Table~\ref{tab:results_summary} summarises the results for the key physical, kinematic, and chemical properties derived for the three sources in this work.

\subsection{Morphology and fragmentation}\label{sec:morphology}

Fig.~\ref{fig:main} presents the moment zero of all methanol lines and the ASHES continuum: CH$_3$OH emission is extended and diffuse in all clumps, appearing as fluffy filaments, while the continuum is more compact.
Flux recovery rates from single-dish to ALMA observations are low. We scale the ATLASGAL 870\,$\mu$m peak flux (corresponding to CoCoA pointing centre) to 1.3\,mm using the full modified blackbody law with dust emissivity index $\beta = 1.5$ and the source-specific dust temperatures from the ATLASGAL catalogue \citep{urquhart_atlasgal_2022}. The ratio of the total ALMA flux (i.e.\ CoCoA ACA+MA) to this scaled single-dish value ranges from $\sim$5\% to $\sim$25\%, rising with $L/M$
(\pillaiOverviewPaper). We define a core formation efficiency (CFE) as the fraction of the scaled single-dish clump flux recovered in dendrogram-identified cores. The CFE ranges from $\sim$6\% to $\sim$23\% (see Table~\ref{tab:results_summary}), closely tracking the overall flux recovery. This indicates that nearly all of the ALMA-recovered emission is concentrated in
compact cores, with negligible diffuse inter-core emission in the continuum images.
The average Jeans mass of the clumps mirrors this trend, decreasing from $3.0\unit{\msun}$ in \gttt\ to $1.8\unit{\msun}$ in \gft\ \citep[computed using Eq.~3 in][]{pillai_probing_2011}.
We used the average density of the clumps as it results from their masses and radii (to be consistent with the area over which the mass was determined in ATLASGAL), assuming they are spherical, and we computed the isothermal sound speed using the dust temperature in Table~\ref{tab:cocoa_sources}.
The corresponding number of thermal Jeans masses contained within the clump (hereafter the Jeans mass count, $M_{cl}/M_J$, shown in Table~\ref{tab:results_summary}) increases with the $L/M$ ratio from $440$ to $1150$. This theoretical capacity for fragmentation directly mirrors the observed increase in the actual fragmentation level of the clumps (i.e., the number of extracted cores).

\begin{table*}[th]
    \centering
    \small
    \setlength{\tabcolsep}{4.5pt}
    \caption{Diagnostics used to discriminate between gravity- and turbulence-driven evolution.}
    \label{tab:results_summary}
    \begin{tabular}{l r r r r r r r r r r}
        \toprule
        Source & $N_c$ & $N_c/A$ & $M_J$ & $M_{cl}/M_J$ & CFE & $n_{cl}(\mathrm{H_2})$ & $\frac{\overline{n_{core}(\mathrm{H_2})}}{\overline{n_{env}(\mathrm{H_2})}}$ & $\chi(\mathrm{CH_3OH})$ & $||\nabla\varv_{long}||$ & $||\nabla\varv_{rad}||$ \\
               &       & ($\mathrm{pc^{-2}}$) & ($\mathrm{\msun}$) & & $\%$ & ($10^4\unit{cm^{-3}}$) & & ($10^{-9}$) & ($\mathrm{km\, s^{-1}\, pc^{-1}}$) & ($\mathrm{km\, s^{-1}\, pc^{-1}}$) \\
        \midrule
        \gttt  & 5  & $4 \pm 2$  & $3.0^{+1.7}_{-1.1}$ & $440^{+1070}_{-310}$  & $5.9 \pm 2.6$  & $4.8^{+4.1}_{-3.0}$  & $1.1 \pm 0.1$ & $2.3 \pm 0.7$ & $0.80 \pm 0.05$ & $3.60 \pm 0.04$ \\
        \gte   & 13 & $8 \pm 2$  & $2.1^{+0.7}_{-0.5}$ & $1070^{+2440}_{-740}$ & $9.8 \pm 2.7$  & $5.3^{+2.3}_{-1.6}$  & $1.2 \pm 0.3$ & $3.8 \pm 1.1$ & $0.685 \pm 0.003$ & $2.955 \pm 0.003$ \\
        \gft   & 23 & $28 \pm 6$ & $1.8^{+0.5}_{-0.4}$ & $1150^{+2560}_{-790}$ & $23.4 \pm 4.9$ & $10.8^{+4.3}_{-2.9}$ & $5.8 \pm 1.2$ & $4.3 \pm 1.3$ & $3.27 \pm 0.02$ & $0.15 \pm 0.02$ \\
        \bottomrule
    \end{tabular}
    \tablefoot{
        The columns indicate the number of cores ($N_c$), the number of cores per unit area ($N_c/A$), the Jeans mass ($M_J$), the Jeans mass count in the clump, the core formation efficiency (CFE), the average density in the clump ($n_{cl}(\mathrm{H_2})$) derived from the methanol fit, the density contrast ($\overline{n_{core}(\mathrm{H_2})}/\overline{n_{env}(\mathrm{H_2})}$), the abundance of CH$_3$OH ($\chi(\mathrm{CH_3OH})$), and the velocity gradients of the selected velocity component in the longitudinal ($||\nabla\varv_{long}||$) and radial direction ($||\nabla\varv_{rad}||$). Uncertainties on core-counting metrics ($N_c/A$, CFE) are derived from Poisson statistics, while abundance errors assume a $30\%$ uncertainty. Parameters heavily dependent on highly uncertain mass or density estimates ($M_J$, $M_{cl}/M_J$, $n_{cl}$) are reported with asymmetric bounds derived via log-space error propagation to ensure positive physical limits. The PPV cubes were rotated counter-clockwise to align the major axis with the $x$ direction by $45.0^\circ$, $-41.4^\circ$, and $45.0^\circ$, respectively.
    }
\end{table*}

Applying a more stringent dendrogram analysis than \citet{sanhueza_alma_2019} and \citet{morii_alma_2024} (with a minimum intensity threshold of $5\sigma$ instead of $3\sigma$, minimum contrast of $1\sigma$, and minimum area equivalent to one beam instead of $0.5$ beams), we identified $41$ cores in the ASHES images.
\gft\ yielded $23$ cores, \gte\ contains $13$, and \gttt\ returned $5$.
All cores are compact and moderately massive (typically $\sim 2 \unit{\msun}$ with radii of $\sim 0.018 \unit{pc}$ or $3700\unit{AU}$), consistent with early-stage fragmentation in high-mass star-forming clumps \citep{sanhueza_alma_2019}.
\gft\ has the highest number of cores per unit area ($N_{c}/A\sim$\,28~cores~pc$^{-2}$), followed by \gte\  ($\sim$\,8~cores~pc$^{-2}$), and \gttt\ has the lowest ($\sim$\, 4~cores~pc$^{-2}$).
This trend reflects the change in the fraction of diffuse material in the clumps, as revealed by the recovered flux.

When comparing the absolute number of detected cores across the sample, it is important to address the effect of distance. G14 is the closest region ($3.1\unit{kpc}$), yielding better spatial resolution and mass sensitivity compared to G332 ($4.0\unit{kpc}$) and G28 ($4.5\unit{kpc}$). To ensure our fragmentation trend is not a distance-driven artifact, we estimated the impact of these limits on the retrieved number of cores.
First, the continuum mass sensitivity scales with the square of the distance. If G14 were at $4.5\unit{kpc}$, the mass sensitivity would degrade by a factor of $(4.5 / 3.1)^2 \approx 2.1$, artificially raising the detection threshold to $\sim 0.74\unit{M_\odot}$. In our catalogue only $4$ of the $23$ cores in G14 fall below this higher threshold.
Second, the minimum physical separation scales linearly with distance, meaning the physical beam size would inflate by $45\%$ at the distance of G28. The minimum separation between any two cores is $\sim11.0$ pixels. Because inflating the synthesized beam by $45\%$ results in a beam width of roughly $\sim9$ pixels along the major axis, the cores remain sufficiently separated such that zero (or at most one) pair would merge due to degraded resolution.
Therefore, even if G14 were observed at the poorest mass sensitivity and spatial resolution of our sample, the resulting core count (18 to 19 cores) would still largely exceed the populations in G28 (13 cores) and G332 (5 cores). We conclude that the rise in the number of cores in our sources is robust against the distance effect.

\subsection{Number density}\label{sec:density_structure}
From the fit results of the methanol lines (Sect.~\ref{sec:obs_methods}) we compute the line ratios that are a direct probe of the average density along the LOS (Sect.~\ref{sec:density_method}), because all the material along that contributes to the integrated intensity \citep{giannetti_ch3oh_2025}.
The density computed for the entire clump increases from $4.8\times 10^4\unit{cm^{-3}}$ in \gttt\, to $5.3\times 10^4\unit{cm^{-3}}$ in \gte, to $1.1\times 10^5\unit{cm^{-3}}$ in \gft\ (cf. Table~\ref{tab:results_summary}).
Our density determinations show that while envelope (see Sect.~\ref{sec:line_fitting}) densities are consistently several $10^4~\mathrm{cm^{-3}}$ (dashed circles in Fig.~\ref{fig:main}), core densities reach up to several $10^5$ cm$^{-3}$ (solid circles in Fig.~\ref{fig:main}, corresponding to continuum or methanol peaks; see Tables~\ref{tab:G332.96_densities} -- \ref{tab:G14_densities}). This indicates a larger density contrast with the envelope in the more fragmented clumps.
Cores associated with active, early SF (identified by outflows and/or warm material in \citet{morii_alma_2024}, marked by white stars in Fig.~\ref{fig:main}) are generally denser than both envelope regions and cores not associated with active SF.
We compute a reliability score for the density fit as a weighted linear combination of the SNR of the lines, their ratio, and its uncertainty. This quantity ranges from $0$ to $1$, with higher values indicating more reliable fits. The exact formula is available in the project's \href{https://www.ict.inaf.it/gitlab/andrea.giannetti/cocoa_methanol_archive/}{repository}.
This reliability score is used as a selection criterion in the computation of the density contrast, where we only use the most robust component for each of the defined regions.
Finally, to compute the average core and envelope volume densities used for the density contrast (Table~\ref{tab:results_summary}), we calculate the inverse-variance-weighted mean of these selected components, using the 67\% HPD interval to define the variance.

To ensure that the higher density contrast observed in \gft\ is not an artifact of spatial averaging over larger envelope apertures (which could artificially lower the inferred density), we performed a control test by re-extracting the envelope spectra in \gft\ and \gte\ using the same aperture size employed for the cores.
The resulting envelope densities remained statistically indistinguishable from those extracted with larger apertures.
These tests confirm that the density of the diffuse envelope is uniform across the sample.
Therefore, the increased density contrast observed in \gte\ and \gft\ (up to a factor of $5-6$) is driven by the higher density of their cores compared to \gttt.

    \begin{figure*}[th]
        \centering
        \includegraphics[width=0.33\linewidth]{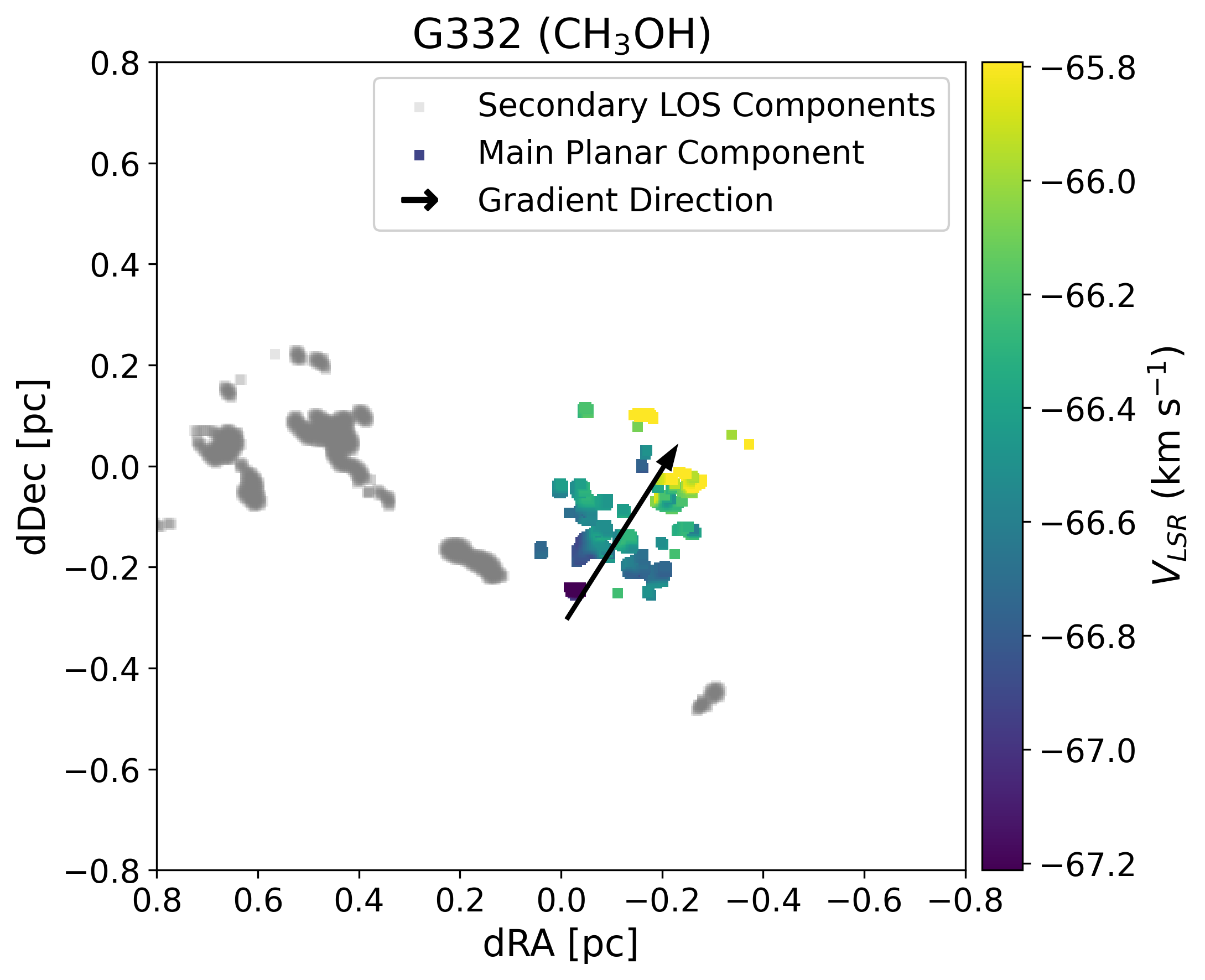}
        \includegraphics[width=0.33\linewidth]{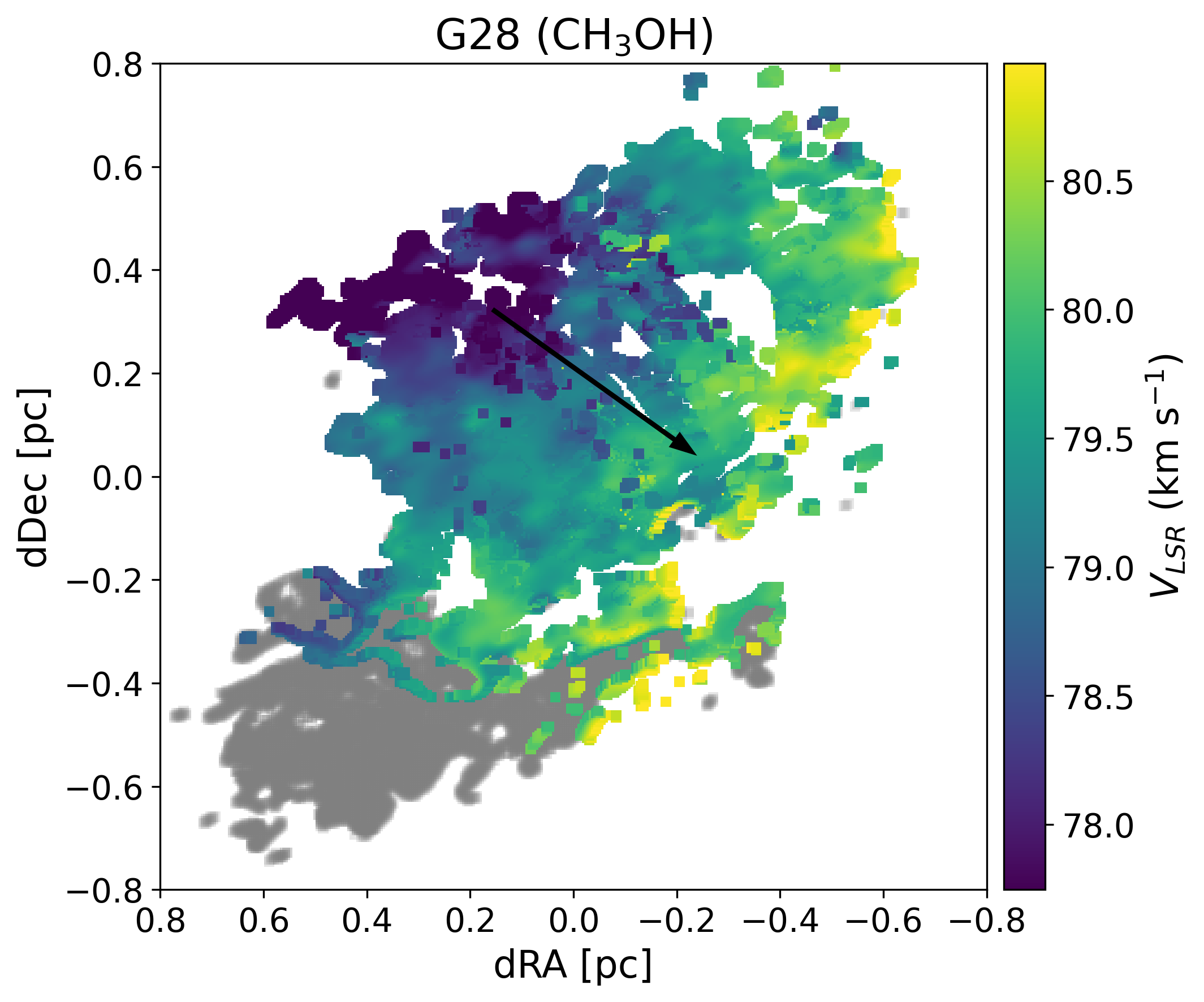}
        \includegraphics[width=0.33\linewidth]{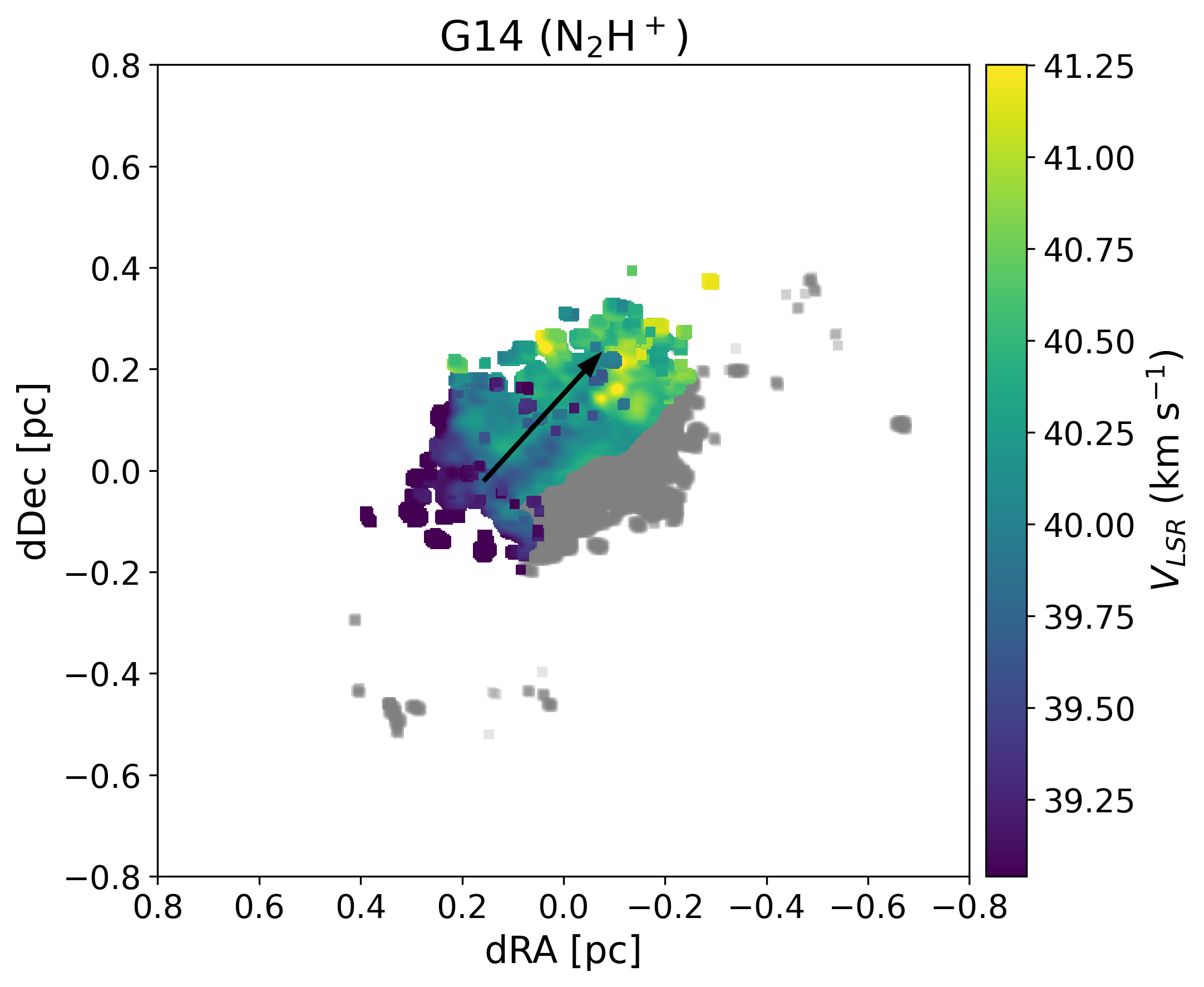}
        \caption{\label{fig:velocity_fields} Projected 2D velocity fields of the sources. The species used is indicated above each panel; both of the species are mutually consistent for velocity measurements. The coloured points display the $V_{\mathrm{LSR}}$ of the velocity component extracted via clustering, which was used for the planar fit. The semi-transparent grey points represent additional, overlapping LOS velocity components that were filtered out before fitting. The black arrows indicate the direction of the fitted planar velocity gradients. An interactive 3D version of the PPV cubes and the fitted planes is provided through a \href{https://colab.research.google.com/drive/1rKI06UAOoi60jMVzRfYv4Wkwhir5Qgq4?usp=sharing}{Colab notebook}.}
    \end{figure*}

\begin{table*}
    \centering
    \caption{Contrasting expectations of fragmentation in massive clumps under the gravity- and turbulence-dominated scenarios.}
    \label{tab:predictions}
    \begin{tabular}{p{3.5cm}p{6.5cm}p{6.5cm}}
        \toprule
        Quantity & Gravity-dominated & Turbulence-dominated \\
        \midrule
        Density (Sect.~\ref{sec:low-fragmentation}) & Increases with time. Lower average densities and less density contrast in less evolved clumps. & Stays roughly constant in the turbulent-support case, or, in general, has no systematic secular evolution if gravity is not dominant. No clear correlation with fragmentation. \\
        Fragmentation (Sect.~\ref{sec:low-fragmentation}) & Triggered by increasing density (decreased Jeans mass). Number of cores increases with clump density and evolution. & Primarily linked to Mach number$^{a}$. Stronger turbulence produces higher-contrast fluctuations.\\
        \raggedright{}Velocity gradients (Sect.~\ref{sec:ordered-motions}) & Ordered motions dominate. Gradients are comparable to linewidths, contributing substantially to them. & Under turbulent-support, clumps are disconnected from the environment, with little to no mass exchange. In the inertial-inflow scenario, gradients are much smaller than linewidths, which are primarily turbulent. \\
        \raggedright{}Core-to-core velocity disp.\ (Sect.~\ref{sec:ordered-motions}) &  Increases with time and fragmentation due to acceleration by gravity. & Increases with the Mach number. \\
        \bottomrule
    \end{tabular}
    \tablefoot{$^{a}$ The number of fragments could increase because the density enhancements become more extreme and the local thermal Jeans mass decreases, or it could decrease, if the effect of turbulence is to increase the support, as suggested by the idea behind the turbulent Jeans mass.}
\end{table*}

\subsection{Kinematics and velocity gradients}\label{sec:kinematics}
From our pixel-by-pixel line-fitting procedure we obtain the cube of velocity centroids of both CH$_3$OH and N$_2$H$^+$.
We compared the centroids obtained from different species and found them to be consistent across all sources.
We observe regions in the PPV space that exhibit well-defined planar gradients.
To quantify these gradients, we isolated these portions of the PPV space using the Bayesian Gaussian Mixture clustering model and fitted a plane to them. This step allows to overcome the complication of having multiple velocity components along the LOS, and helps to select a coherent structure.
The gradients of the most prominent structures are shown in Figure~\ref{fig:velocity_fields}.

Throughout this work, we interpret the observed continuous velocity gradients along and across the principal axes of the clumps as signatures of mass flow and infall, as is common in the literature for this type of objects \citep[e.g.][]{peretto_global_2013, kirk_filamentary_2013, henshaw_dynamical_2014, motte_high-mass_2018, morii_global_2025}. However, we note that large-scale velocity gradients can also originate from other scenarios, such as rotation, expansion, or the line-of-sight superposition of distinct clouds.
Given the early evolutionary stage of our sample, widespread clump-scale expansion driven by feedback is physically unlikely.
The risk of misinterpreting the superposition of two distinct clouds as a continuous velocity gradient across the clump is also mitigated by our pixel-by-pixel multi-component spectral fitting. Nevertheless, we cannot completely rule out the contribution of bulk rotation, or clouds very close to each other in velocity space and spatially coherent.
Therefore, our infall metrics should be viewed as tracing the most likely, but perhaps not exclusive, kinematic mechanisms generating the gradients in these regions.

Subsequently, we rotated the fitted plane to align the major axis of the "filaments" with the $x$-axis.
This rotation allows the plane parametrisation to directly provide the radial velocity gradient components along and across the filamentary structure.
For \gte\ and \gttt, we used CH$_3$OH emission to quantify the velocity gradient, which was stronger and more extended than that of the isolated component of N$_2$H$^+$.
For \gft, we instead used N$_2$H$^+$ because its velocity field exhibited clearer structure, facilitating more robust clustering and fitting.
All three sources show large velocity gradients (up to $\sim 3.5~\mathrm{km \, s^{-1} \, pc^{-1}}$; Fig.~\ref{fig:velocity_fields}), comparable to the observed single-dish linewidths.
In the less fragmented sources (\gttt\ and \gte), radial motions directed towards the filamentary structures dominate the velocity field. In contrast, the more fragmented source, \gft, shows more prominent longitudinal motions along its major axis.

    \begin{figure*}[th]
        \sidecaption
        \includegraphics[width=1.43\columnwidth]{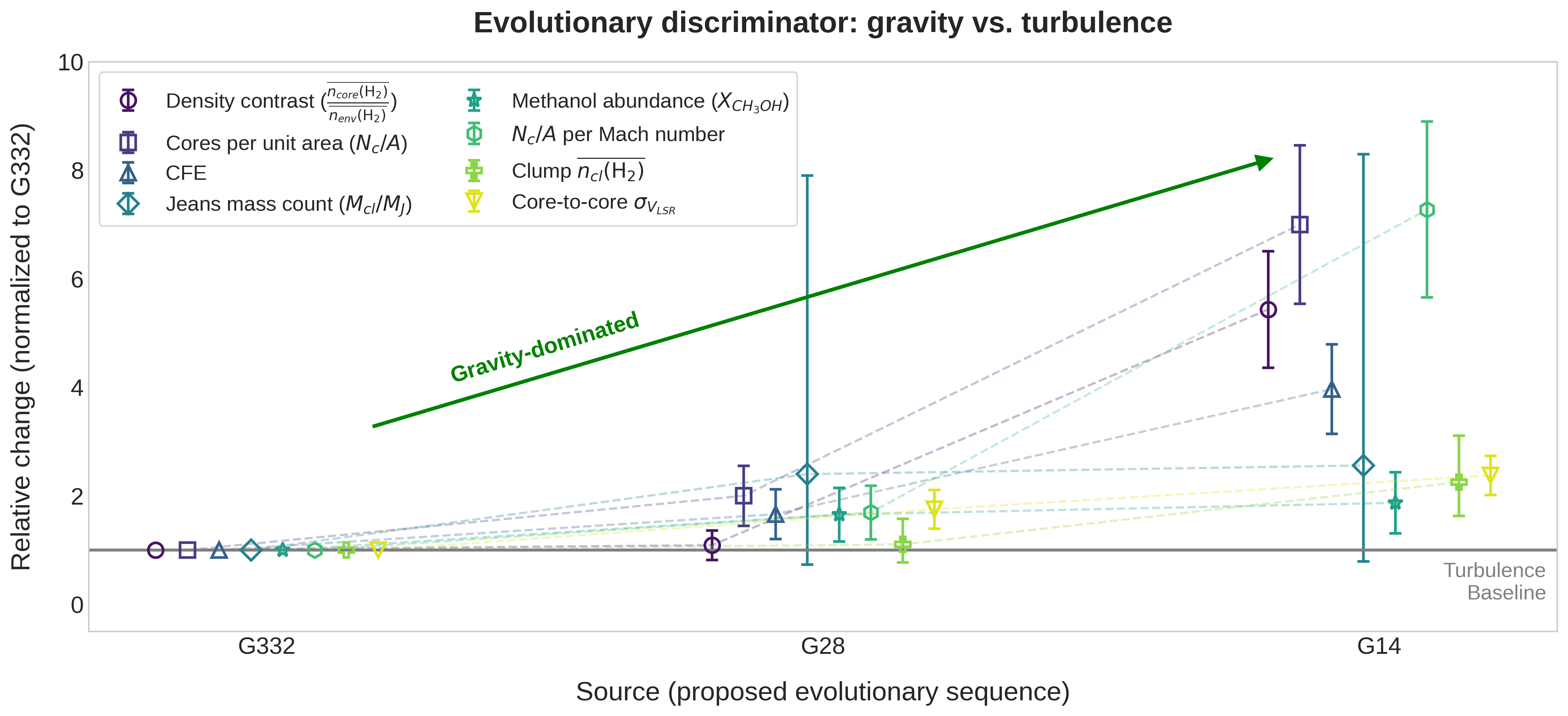}
        \caption{Relative change of key physical quantities with the proposed evolutionary sequence. All quantities are normalised to the least-evolved source, \gttt, establishing the horizontal grey line ($y=1$) as the baseline for no change, the expected trend for a turbulence-driven scenario where core and clump parameters remain relatively constant. The green arrow is a qualitative representation indicating that these parameters are expected to increase over time in a gravity-dominated scenario; it does not correspond to a specific numerical rate of increase. Error bars represent the propagated uncertainties (see Table~\ref{tab:results_summary}).}\label{fig:trends}
    \end{figure*}

To estimate the filamentary infall rate ($\dot{M}$) onto \gte, we applied the model detailed in $\text{Section \ref{sec:infall_method}}$.
It assumes projected dimensions of $L_p \approx 1 \unit{pc}$ (major axis) and $W_p \approx 0.5 \unit{pc}$ (minor axis), derived from the spatial extent of our moment-zero maps (Fig.~\ref{fig:main}, middle panels).
We estimate that the intrinsic thickness, $H$, is around $0.1 \unit{pc}$.
The path length along the LOS is approximately $0.15 - 0.2 \unit{pc}$, obtained by dividing the clump's $\mathrm{H}_2$ column density from the ATLASGAL dust continuum by the mean clump density reported in Table~\ref{tab:results_summary} (cf.~Eq.~\ref{eq:los_depth}). Accounting for the effect of inclination ($H = H_{\mathrm{LOS}} \cos \gamma \cos \theta$; see Appendix~\ref{app:geometry}), this yields an intrinsic thickness of $\approx 0.1 \unit{pc}$.
The resulting sheet-like morphology is consistent with the presence of linear infall gradients onto filaments \citep{chen_self-gravitating_2020}.
Using its measured radial velocity gradient ($||\nabla\varv_{rad}|| \approx 2.96 \unit{km \, s^{-1} \, pc^{-1}}$ from Table~\ref{tab:results_summary}) and the clump average density of $5.3\times 10^4\unit{cm^{-3}}$, we calculate an order-of-magnitude estimate for the total filamentary infall rate:
\begin{equation}
    \dot{M}_{\mathrm{fil}} \approx 730 \unit{\msun\, Myr^{-1}} = 7.3 \times 10^{-4}\unit{\msun\, yr^{-1}}.
    \label{eq:infall}
\end{equation}
For \gttt\ we assume $L_p \approx 0.7 \unit{pc}$ and $W_p \approx 0.2 \unit{pc}$; also in this case, the LOS dimension, $H$, results to be $\sim0.1 \unit{pc}$.
The radial velocity gradient is $||\nabla\varv_{rad}|| \approx 3.6 \unit{km \, s^{-1} \, pc^{-1}}$ for this source, and the mean clump density is $4.8\times 10^4\unit{cm^{-3}}$.
The filamentary infall rate onto \gttt\ is then $\dot{M}_{\mathrm{fil}} \approx 220 \unit{\msun\, Myr^{-1}} = 2.2 \times 10^{-4}\unit{\msun\, yr^{-1}}$.
This high infall rate is comparable to single-dish estimates for other high-mass star-forming regions \citep{he_infall_2015, wyrowski_infall_2016, traficante_testing_2018, pillai_infall_2023}, suggesting that the large-scale flow in the envelope still actively supplies mass to the central high-density structures.
While both numbers are order of magnitude estimates, we note that the mass delivered onto the filament would increase with evolution.

\subsection{Chemical properties}
The offset between the continuum and molecular peaks Figure~\ref{fig:main} could indicate that methanol and N$_2$H$^+$ are freezing out in the densest parts of the cores, hindering our ability to trace them and obtain a reliable estimate of $n(\mathrm{H_2})$. For that to happen, densities need to exceed $10^6\unit{cm^{-3}}$, the threshold where N$_2$H$^+$ depletes \citep{pillai_probing_2011}. Indeed, densities from the ASHES continuum data indicate they can get that high \citep{morii_alma_2023} in cores. But cores are not isolated structures; they're embedded in lower-density envelopes. For us to see no increase in the LOS-averaged density, the core's contribution to the line emission would have to be negligible to hide density enhancements as low as a factor of two.

How depletion works is not straightforward. Naively, one could think that once a protostar forms and starts heating up the surrounding material CO depletion would decrease. \citet{sabatini_alma_2022} showed the opposite: depletion is actually higher in young protostellar cores than quiescent ones because their envelopes remain cold and dense. This might be why some actively star-forming cores have molecular gas densities that look more like the surrounding envelope, and lower than what we see in a typical core in our data.
Another possibility to explain the low densities measured in some cores is that some continuum peaks without active SF are not true cores at all, but just chance superpositions of material along the LOS, as \citet{padoan_will_2023} suggested. Some indications suggest that multiple velocity components exist at the location of the central cores in \gte.
\gft\ cores always show larger densities, which may indicate a more advanced SF activity: if the embedded objects are sufficiently luminous to sublimate the ices on grains in a significant portion of the object, methanol can trace the densest layers of the core, revealing clear overdensities along the LOS.

Finally, we calculated the clump-averaged methanol abundance by computing the methanol column density over the ATLASGAL beam aperture and dividing it by the peak $\mathrm{H_2}$ column density from ATLASGAL.
It increases slightly with fragmentation, from $2.3 \times 10^{-9}$ in \gttt\ to $4.3 \times 10^{-9}$ in \gft. The value for \gte\ is $3.8 \times 10^{-9}$, which agrees well with the value reported in \citet{sanhueza_distinct_2013}.

\section{Discussion}\label{sec:discussion}

The fundamental question regarding the minimally fragmented clumps in the CoCoA sample is whether their lack of substructure is a permanent feature generated by anomalously low turbulent initial conditions, or if it represents an extremely early, transient evolutionary stage prior to the significant formation of cores.
While previous dust continuum surveys have established a statistical correlation between lower fragmentation levels (or continuum substructure) and younger evolutionary indicators across large samples \citep[via the $L/M$ ratio and dust temperature][]{elia_almagal_2026, wallace_almagal_2026}, individual sources at these early stages exhibit significant scatter, meaning that the degree of continuum substructure alone cannot unambiguously establish the evolutionary stage of a specific source.

To break the degeneracy between static initial properties and dynamic evolution, our study uses an approach built upon three new and independent elements:
\begin{enumerate}
    \item Direct gas-phase volume density estimates: Using the newly calibrated CH$_3$OH line ratios, we measure the number density on ALMA scales independently of dust properties, temperature, geometry, or assumed gas-to-dust ratio;
    \item Unfiltered kinematic data: By combining all ALMA arrays, we recover the full velocity field, allowing us to trace both large-scale inflows and local core-to-core velocity dispersions without interferometric filtering;
    \item Astrochemical timescales: Our measured gas-phase CH$_3$OH abundances provide independent constraints on the duration of this pre-core phase that can be compared to the mass build-up time of the clumps.
\end{enumerate}

By applying these diagnostics to our sample, we characterise the systematic variations in the gas phase as we move from the diffuse, minimally fragmented clumps (\gttt\ and \gte) to the highly structured environment of \gft. Turbulence-supported and gravity-dominated scenarios make contrasting predictions for how these physical and chemical properties should co-evolve during internal clump fragmentation, as summarised in Table~\ref{tab:predictions}. We therefore assess which of these theoretical frameworks can better reproduce this observed combination of gas density, kinematics, and abundance trends.

\subsection{Low fragmentation is linked to a lower clump density\label{sec:low-fragmentation}}
    Within our sample, the clumps exhibiting a minimal level of fragmentation (\gttt\ and \gte) are characterised by dominant extended emission and low continuum flux recovery. As a preliminary statistical indicator of an evolutionary progression, the $L/M$ ratio increases alongside the fraction of recovered continuum flux and CFE across the three sources, suggesting that a larger portion of the gas is progressively funneled into compact structures as clump evolution advances \citep[in agreement with][]{sanhueza_alma_2019, elia_almagal_2026}.

    This proposed sequence aligns with broader statistical trends in dust emission, where structural complexity correlates with average clump density and temperature throughout the early cluster formation timeline \citep{zari_herschel-planck_2016, alfaro_primordial_2018, palau_does_2021, xu_alma_2024, morii_alma_2024, wallace_almagal_2026}.
    Our direct measurements of $n(\mathrm{H_2})$ via CH$_3$OH line ratios allow us to physically test this evolutionary hypothesis in the gas phase. We observe a clear, systematic increase in both the average volume density and the core-to-envelope density contrast from the diffuse to the structured sources (Table~\ref{tab:results_summary}), independently confirming a dynamic structural evolution without relying on uncertain dust models, geometric or statistical assumptions alone.

    These results are consistent with the GHC scenario \citep{vazquez-semadeni_global_2019}. As clumps evolve and accrete more material, they become denser, their average Jeans mass decreases, permitting the formation of progressively more and denser cores, and triggering the top-down process of fragmentation.
    Because collapse timescales significantly shorten with density, the core-to-envelope density contrast increases with evolution \citep{camacho_simultaneous_2020}.

    On the contrary, in scenarios that rely on turbulence to provide support against gravity, the clumps are maintained in a quasi-steady state \citep[e.g., ][]{zuckerman_models_1974, mac_low_control_2004, mckee_theory_2007}.
    In this framework, turbulent support prevents the global contraction that would compress the clump to higher mean densities. Therefore, there is no physical mechanism driving a progressive and significant increase in average density, in contrast to the gravity-dominated scenarios of clump-scale fragmentation \citep{vazquez-semadeni_turbulent_2026}. Furthermore, in turbulent support models, the relative core-to-envelope density contrast is determined by the isothermal shock compression ratio, which scales with the Mach number squared \citep{mac_low_control_2004}.
    Had turbulence been the primary mechanism causing the large observed linewidths ($\approx 3-4 \unit{km\,s^{-1}}$, measured from methanol spectra averaged over the entire ALMA primary beam to capture the full clump scale), the resulting Mach numbers ($\mathcal{M} \approx 6-8$) would be similar across our sample.
    As a consequence, a purely turbulence-regulated fragmentation scenario predicts a roughly constant density contrast and similar fragmentation levels for all three clumps \citep[cf.][]{guszejnov_isothermal_2018}. Instead, we observe that the core-to-envelope ratio increases by a factor of $5-6$, and the clump average density by a factor of $\approx 2$ (Table~\ref{tab:results_summary}) from the minimally-fragmented sources to the highly-fragmented one, challenging a turbulent support-dominated state.
    This progressive increase in the clump average density and in the density contrast rules out initial conditions as the cause of the differences among the sources, and confirms that these massive clumps are actively evolving within a gravity-dominated regime, rather than being maintained in a quasi-static, turbulence-supported state \citep{vazquez-semadeni_turbulent_2026}.
    A caveat to this Mach-number scaling is that the turbulent driving mode (solenoidal versus compressive) also influences density fluctuations and fragmentation, as more compressive driving produces higher density contrasts and denser fragments \citep{federrath_comparing_2010}. However, our minimally fragmented sources show both a lower degree of fragmentation and lower core densities, a combination that still cannot be reconciled with a higher compressive fraction.

    \subsection{Ordered motions dominate over turbulence\label{sec:ordered-motions}}
    The kinematics provide another crucial signature of this process.
    When evaluating these motions, it is helpful to distinguish between the mechanism that initially assembled the clumps and the force currently driving their internal evolution.
    In the inertial-inflow scenario, ordered inflow velocities are predicted to be much lower than the internal turbulent velocity dispersion, as measured by the clump linewidths \citep{padoan_origin_2020}. On the contrary, if the clump is currently undergoing global collapse, infall velocities must constitute an important fraction of the observed linewidths \citep{vazquez-semadeni_turbulent_2026}.
    Our sources show large velocity gradients that are comparable to the line FWHMs integrated over the full primary beam, covering the entire ATLASGAL clumps.
    This indicates that systematic, ordered motions, rather than turbulent, disordered ones, now dominate the linewidth at pc-scales.

    Indeed, the estimated gravitational velocity ($\approx\sqrt{2 (G\,M) / R}$, where $M$ is the mass of the clump given in Table~\ref{tab:cocoa_sources}) ranges from $3.4$ to $4.4 \unit{km\,s^{-1}}$ at $1\unit{pc}$.
    This is within $40\%$ of the velocity differences produced by the observed gradients across the same scale, especially considering that the latter are computed using only the projected radial component of the velocity field.
    While these ordered, pc-scale motions could initially arise from turbulent converging flows responsible for assembling these structures \citep{padoan_origin_2020}, their close agreement with the gravitational velocity indicates that gravity has since become the dominant driving force regulating the internal dynamics of these clumps \citep{vazquez-semadeni_global_2019, vazquez-semadeni_turbulent_2026}.

    A significant contribution from large-scale ordered motions to the total linewidth has also been reported in more evolved samples, where the two-point correlation function of the velocity field reveals that significant linewidths are measured on large scales \citep[Table~4 of][]{palau_does_2021}.
    Overall, these ordered flows are the mechanism at the root of the increased mass of the most massive core and morphological complexity with evolution observed in CoCoA (\pillaiOverviewPaper) and ALMAGAL \citep{elia_almagal_2026, wallace_almagal_2026}.

    This dominance of gravity at the clump scale agrees with recent observational and numerical work.
    All three clumps in our sample have high surface densities ($\Sigma \approx 0.5 - 2.0 \unit{g\,cm^{-2}}$), significantly exceeding the critical threshold of $\Sigma \simeq 0.1 \unit{g\,cm^{-2}}$ ($\approx10^4\unit{cm^{-3}}$) identified by \citet{traficante_multiscale_2020}, above which clump dynamics are observed to become gravity-driven.
    We show that this gravitational dominance is not only limited to the dynamics of clumps, but it regulates directly their initial fragmentation process as well.
    Furthermore, numerical simulations tracking gas dynamics across multiple scales find that while large-scale cloud assembly is initially driven by turbulence, gravity becomes the dominant force as mass accumulates and the gas becomes molecular \citep[e.g., ][]{ibanez-mejia_gravity_2022, brucy_what_2025}.
    Also in these frameworks, gravity governs the global dynamics at the clump scale ($n(\mathrm{H_2})\approx 10^4\unit{cm^{-3}}$), while localised fragmentation into cores happens at higher densities ($n(\mathrm{H_2})\approx 10^5\unit{cm^{-3}}$) \citep{appel_what_2023}, in rough agreement with what we observe.
    By accurately determining the gas-phase number density and unfiltered kinematics right before widespread core formation takes hold, we provide the first direct observational evidence of gravity actively driving uniform clumps toward significant fragmentation.

    Though not strictly classical filaments, our fluffy, filamentary clumps may behave analogously to their larger-scale counterparts.
    Filaments are considered the primary suppliers of gas from the environment to the denser clumps where conditions become conducive for the formation of stellar clusters.
    They are not mere conveyor belts moving mass only along their longest axis.
    Rather, they function like highways, collecting traffic from surrounding areas and concentrating it towards major centres, the clumps.
    The development of significant longitudinal infall motions in the more fragmented source (see Table~\ref{tab:results_summary}) is consistent with simulations of gravitational collapse in filamentary structures \citep{gomez_filaments_2014}, where radial accretion builds up mass before longitudinal flows emerge.
    The GLASHES pilot study \citep{morii_global_2025} detected a longitudinal gradient in another filamentary source that is between \gte\ and \gft\ in terms of evolution ($L/M$ ratio and protostellar core fraction), further supporting this evolutionary scenario.
    Linear gradients are also consistent with sheet-to-filament collapse simulations, contrary to isotropic accretion onto filaments \citep{chen_self-gravitating_2020}, as mentioned in Sect.~\ref{sec:kinematics}.

    The filamentary infall rates for \gte\ and \gttt\ (Sect.~\ref{sec:kinematics}) suggest a clump mass accumulation timescale of a few million years, also given that the filamentary infall rate increases with time in very early phases \citep{gomez_filaments_2014, naranjo-romero_gravity-driven_2022}, as our sources also seem to suggest. Because these sources are only marginally fragmented, this represents a rough upper limit for the pre-core phase duration.
    On the other hand, considering the timescale of the cluster formation process after the clumps have accumulated sufficient mass to form high-mass stars \citep[$1-5\times10^5\unit{yr}$, e.g.][]{sabatini_establishing_2021, urquhart_atlasgal_2022}, the total mass added to clumps from Quiescent to UCHII stage would amount to a few hundred solar masses at most, a relatively small fraction of the material in the clump.
    Hints of such a small increase can be seen in Fig.~5 of \citet{wallace_almagal_2026}, where clumps with rich substructure appear to be slightly more massive than those with simpler emission hierarchies, on average.
    The mass accumulation timescales that we infer are roughly consistent with the time it takes to form a massive star, capable of ionising the surrounding medium, and they are comparable to the timescale to form a $\sim10\unit{\msun}$ star in the multiscale accretion model of \citet{vazquez-semadeni_multiscale_2024}, providing additional support to the gravitational collapse scenario.
    The comparison between the pre-core phase lifetime and that of high-mass clumps indicates that cluster formation speeds up with time, as predicted if gravity drives the collapse \citep{vazquez-semadeni_global_2019}.

    The kinematics of the dense cores provide an additional test for these frameworks.
    In gravity-driven scenarios, global contraction and mass accretion deepen the overall gravitational potential well of the clump.
    As the clump evolves, the cores fall inward \citep[cf. also][for the decrease of the distance between cores]{xu_alma_2024} and are accelerated by gravity.
    This trend is a key prediction of the GHC model, where the cores, formed hierarchically, are dynamically coupled to the gravitational potential of the parent clump.
    Indeed, the $V_{LSR}$ dispersion for the extracted core positions (see Tables~\ref{tab:G332.96_densities}-\ref{tab:G14_densities}) increases from the least-evolved ($\sim 0.8\unit{km\,s^{-1}}$ in \gttt) to the most-evolved ($\sim 1.4\unit{km\,s^{-1}}$ in \gte, and $\sim 1.9\unit{km\,s^{-1}}$ in \gft).
    This increasing internal velocity dispersion suggests that the clumps are becoming a dynamically active cluster, with their motions influenced by the deepening, gravitational potential, consistent with the overall acceleration of the SF process.
    By moving faster, the cores may also increase the amount of mass that they accrete from the environment.

\subsection{High methanol abundance suggests a long pre-core phase}
Methanol chemistry independently supports this long pre-core phase.
Observations revealed large quantities of CH$_3$OH and even more complex species in the envelopes of low-mass prestellar cores \citep[][]{bizzocchi_deuterated_2014, scibelli_prevalence_2020}.
The high abundance of complex molecules in such a cold, dense environment triggered the exploration of efficient mechanisms to release them into the gas phase \citep[e.g.][]{vasyunin_formation_2017, garrod_formation_2022}.
Our results extend this finding to an entirely different scale, suggesting that these mechanisms are important on parsec scales in high-mass clumps.
\citet{priestley_neath_2025} offer one potential explanation, finding that complex organic molecules begin to form before cores appear, and are returned to the gas phase via desorption due to H$_2$ formation, cosmic-ray and UV-photon impacts.
The high overall gas-phase abundances of methanol in the coldest, least-evolved sources (such as those in ASHES and CoCoA) could arise from spending several million years at intermediate densities ($\mathrm{few} \times 10^{3-4}\unit{cm^{-3}}$, close to our observed envelope densities) before cores appear, seeding them with complex species.
This timescale agrees with the mass accumulation timescale derived from our kinematic data, supporting our interpretation of an evolutionary sequence driven by gravitational collapse and mass accretion.

\subsection{Magnetic fields delay collapse preserving pristine stages}
Magnetic fields are often invoked to explain the widely different fragmentation of massive clumps, although their precise role is still debated \citep[e.g.,][]{commercon_collapse_2011, pillai2015, palau_does_2021, beuther_density_2024, klos_role_2025}. While the number of cores correlates significantly more with the clump density than the mass-to-flux ratio \citep{palau_does_2021}, \citet{klos_role_2025} suggest that strong magnetic fields can lead to less filamentary and substructured clumps in the early collapse, but the thermal Jeans mass still determines the fragmentation properties.
However, typical magnetic fields generally only delay, rather than stop, the collapse \citep{beuther_density_2024, klos_role_2025}.
This implies that, despite the lack of magnetic field estimates in our sources, if \gttt\ and \gte\ were more magnetically supported than \gft, they would still represent very early, more pristine stages of collapse \citep{klos_role_2025}.

\subsection{Gravity-driven evolution in clump fragmentation}\label{sec:gravity_dominates}

The evolutionary framework suggested by our data, from the more quiescent \gttt\ and \gte\ to the fragmented \gft, provides compelling evidence that massive clumps form an evolutionary sequence driven by gravity. This evolution is consistently tracked by multiple diagnostics discussed above (cf. also Table~\ref{tab:predictions}) and summarised in Fig.~\ref{fig:trends}:
\begin{itemize}
    \item The direct link between increasing average density and enhanced fragmentation, core-envelope density contrast, Jeans masses, and their count supports the GHC scenario;
    \item The detected ordered motions account for a significant fraction of the observed linewidth at clump scale and are comparable to the gravitational velocity, challenging predictions for purely turbulence-driven fragmentation and reinforcing the connection between density and evolution;
    \item The core-to-core velocity dispersion increases with the fragmentation level, as expected when gravity drives the collapse and dynamically accelerates the cores \citep{vazquez-semadeni_global_2019};
    \item The high abundance of methanol in all clumps requires them to spend millions of years in a regime of intermediate densities \citep{priestley_neath_2025}, a timescale that agrees well with the clump mass accumulation timescale derived from our estimates of filamentary infall.
\end{itemize}

In summary, all our findings consistently support the idea that these sources form an evolutionary sequence, revealing an under-represented and very early phase of massive SF.
The minimally-fragmented clumps represent the initial conditions for the hierarchical process that ultimately forms stellar clusters, where large-scale gravitational infall builds up mass and chemical complexity before fragmentation begins.
These results based on a statistically-limited sample constitute excellent tests to apply to larger samples.

\section{Conclusions}
While our analysis focuses on a modest sample of three massive clumps, they span a critical morphological transition: from minimal fragmentation to a highly-fragmented state.
Our analysis of this sample shows that CoCoA sources with low fragmentation are predominantly diffuse and extended, with a lower average density than their highly fragmented counterparts.
This is confirmed by multiple independent indicators:
\begin{itemize}
    \item High fractions of missing continuum flux in the ACA+MA data, caused by interferometric filtering of their uniform, extended emission;
    \item A low core number per $\mathrm{pc}^2$ in both continuum and line emission, confirming the lack of compact, dense substructures;
    \item A more uniform density distribution and low core densities ($\sim 10^5\unit{cm^{-3}}$), which are 3 to 10 times lower than those typically found in the more evolved, fragmented region.
\end{itemize}
Within this limited sample, these findings consistently demonstrate an increasing fraction of dense gas in the fragmented clump, supporting the idea that the degree of fragmentation tracks the progress of gravitational collapse \citep[e.g.][]{vazquez-semadeni_molecular_2018, camacho_simultaneous_2020, klos_role_2025}.
The clumps with minimal fragmentation are therefore likely to be in an early evolutionary phase, likely at the onset of core formation.
Based on the filamentary infall rates estimated for the two least-evolved sources (Sect.~\ref{sec:kinematics}) and the methanol abundance, we suggest this evolutionary stage may last up to a few million years.

Although based on a small number of sources, our results (the existence of a relatively long pre-core phase, the increasing number of cores per $\mathrm{pc}^2$, the increasing fragmentation, and the rising SF rate with increasing density), summarised in Fig.~\ref{fig:trends}, are all consistent with the GHC scenario, indicating that the clumps are evolving within a gravity-dominated regime \citep[e.g.][]{vazquez-semadeni_global_2019, vazquez-semadeni_turbulent_2026}, which regulates their fragmentation.
On the other hand, the lack of correlation between Mach number and fragmentation level or core-to-envelope density contrast (Sect.~\ref{sec:low-fragmentation}) rules out turbulent initial conditions as the root cause of the different morphologies and core properties.
The kinematics of the sample reinforces this conclusion:
\begin{itemize}
    \item Ordered motions dominate over turbulence, with large-scale velocity gradients accounting for a significant fraction of the clump-scale linewidths, which is also comparable to the gravitational velocity in the clumps. These ordered flows build up the mass of the clumps and their structural complexity;
    \item The core-to-core velocity dispersion increases with density and fragmentation, reflecting the dynamical acceleration of cores within a deepening gravitational potential.
\end{itemize}
Finally, the theoretical prediction that the correlation between core number and clump density strengthens over time is supported by comparing these CoCoA and ASHES sources \citep{morii_alma_2024} with more evolved regions \citep{palau_fragmentation_2014, xu_alma_2024}, as well as within the larger ALMAGAL sample \citep{elia_almagal_2026, wallace_almagal_2026}, lending additional support to the GHC framework.
We emphasise that while the close agreement between pc-scale velocity gradients and linewidths suggests gravity as the dominant driver of current clump dynamics and internal fragmentation, our data remain agnostic to whether these structures were initially assembled by larger-scale turbulent compressions.

Our results suggest that massive clumps transition from a diffuse, minimally fragmented state to a dense, highly structured one under the influence of sequential global gravitational collapse.
Extrapolating this finding to the most extreme CoCoA sources that show no fragmentation in the survey data, we propose that they likely represent the true, pristine initial conditions of clump evolution.
Our pilot study of the transition out of this phase provides a unique opportunity to identify and characterise the onset of the cluster formation process, which generates the vast majority of stars in the Universe.

\begin{acknowledgements}
We wish to thank both the anonymous referees (for the theoretical and observational parts of the paper) for contributing to the clarity and rigour of this work.
This work benefited from the UNAM-NRAO Memorandum of Understanding in the framework of the ngVLA Project (MOU-UNAM-NRAO-2023).
P.S. was partially supported by a Grant-in-Aid for Scientific Research (KAKENHI Number JP26H02066) of JSPS. A.P. acknowledges financial support from the UNAM-PAPIIT IN120226 grant, and the Sistema Nacional de Investigadores of References SECIHTI, M\'exico.
A.G. wishes to thank F. Capasso for the deep and inspiring conversations and the continued support over the development of this work.
\end{acknowledgements}

\bibliography{references.bib}

\onecolumn
\begin{appendix}

\section{Validation of interferometric data combination}\label{app:feathering_verification}

Because our analysis heavily relies on measuring the unfiltered kinematics and total gas volume densities of pc-clumps, it is crucial to ensure that the interferometric data combination process is accurate, and it preserves the total flux as seen by the single-dish.

To verify the accuracy of the feathered cubes, we compared the mean brightness temperature over a large circular aperture ($R = 40\arcsec$) centred on the sources. This large aperture ensures that the broad spatial wings of the single-dish beam are included, allowing for a one-to-one comparison of the surface brightness.

\begin{figure*}[!ht]
    \centering
    \includegraphics[width=0.41\linewidth]{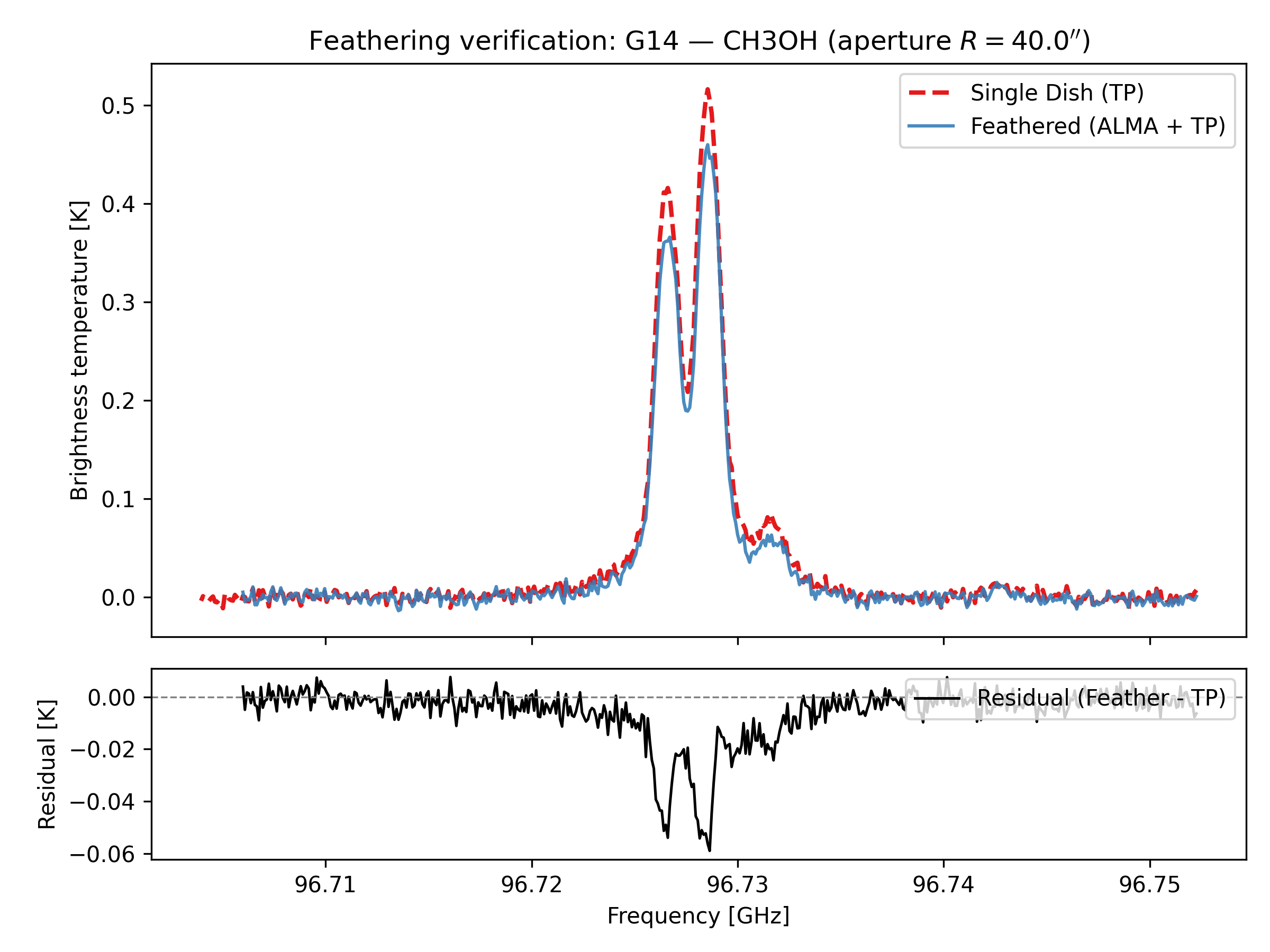}
    \hspace{0.1\linewidth}
    \includegraphics[width=0.41\linewidth]{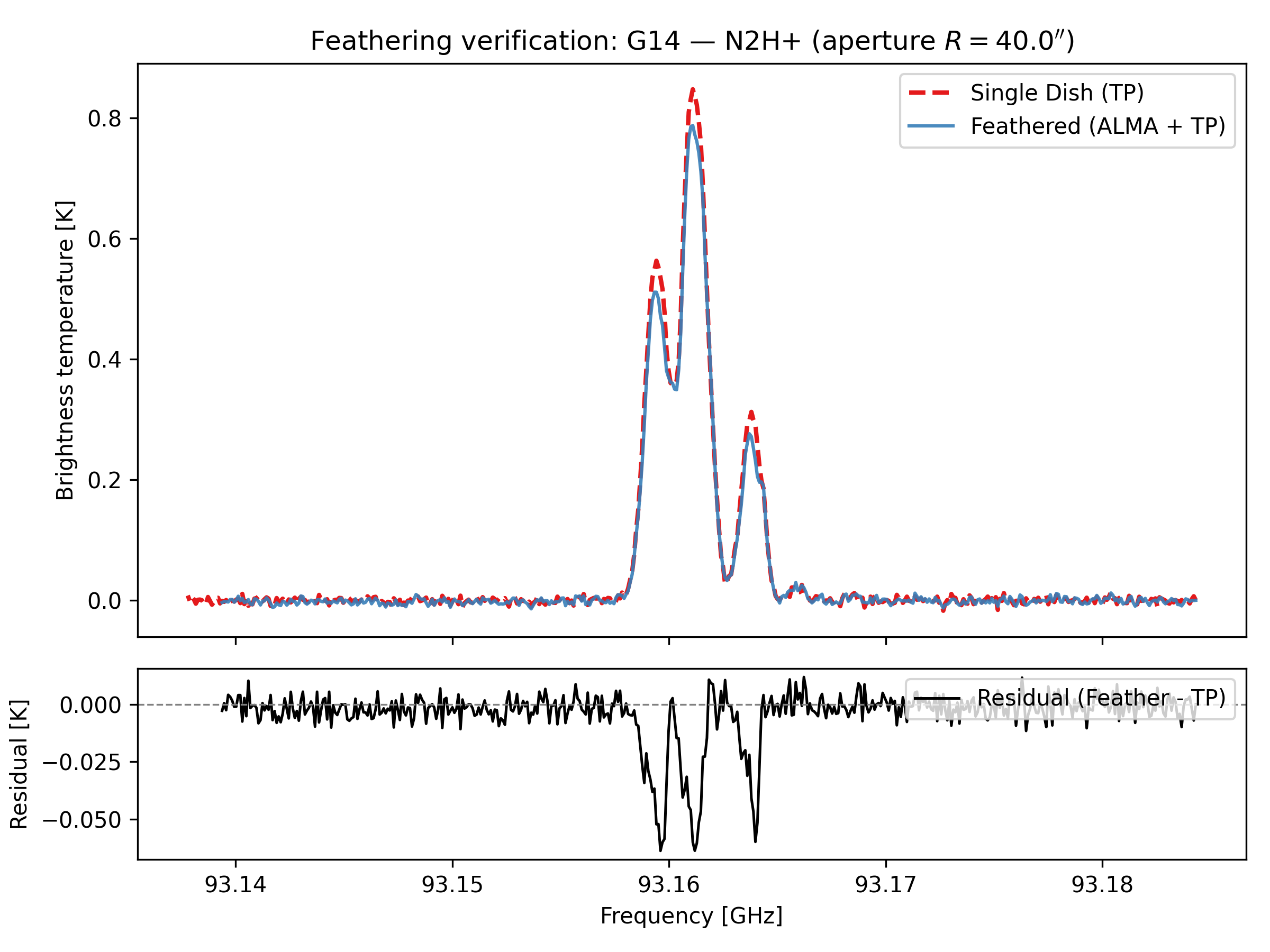}
    \includegraphics[width=0.41\linewidth]{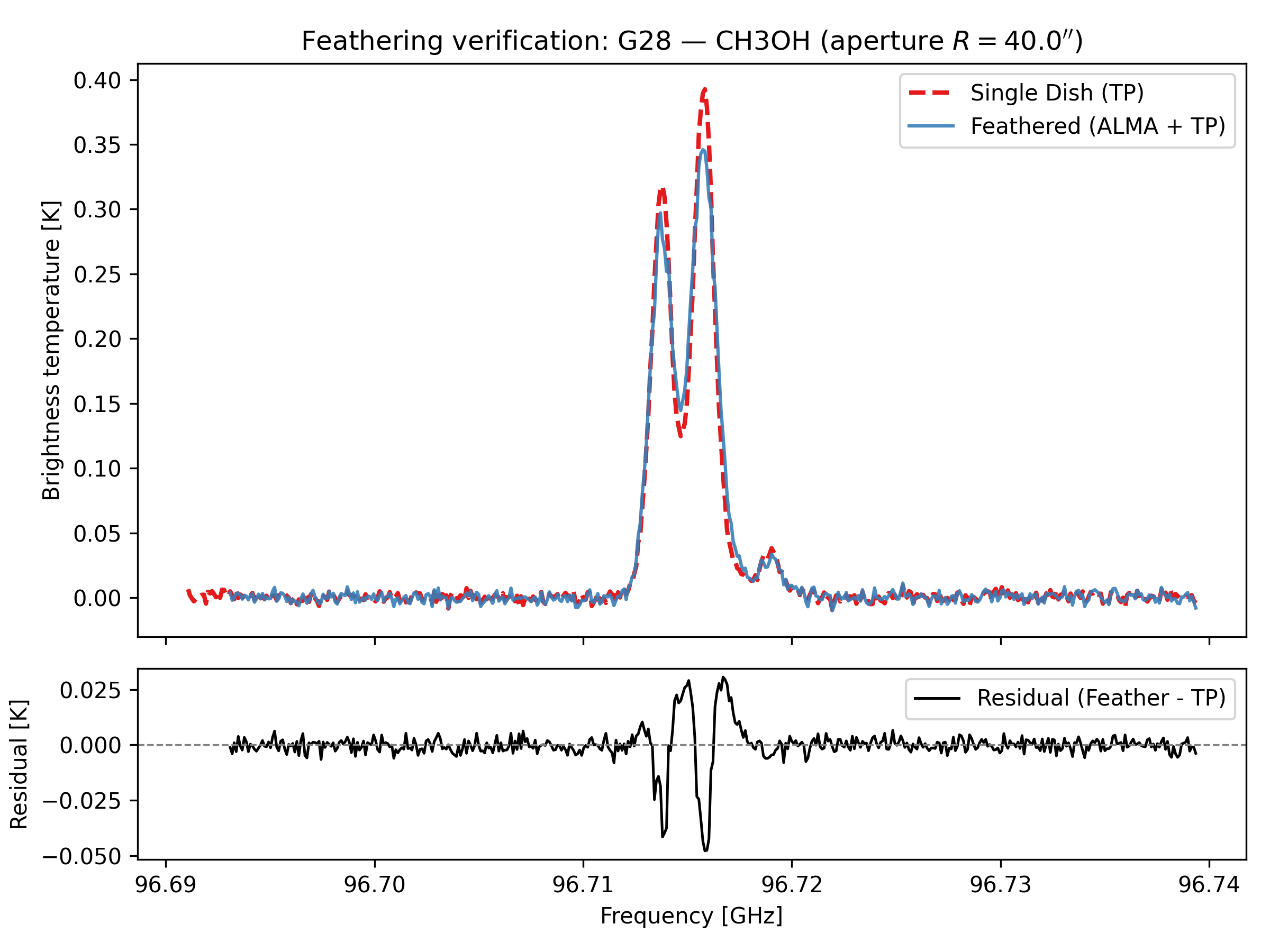}
    \hspace{0.1\linewidth}
    \includegraphics[width=0.41\linewidth]{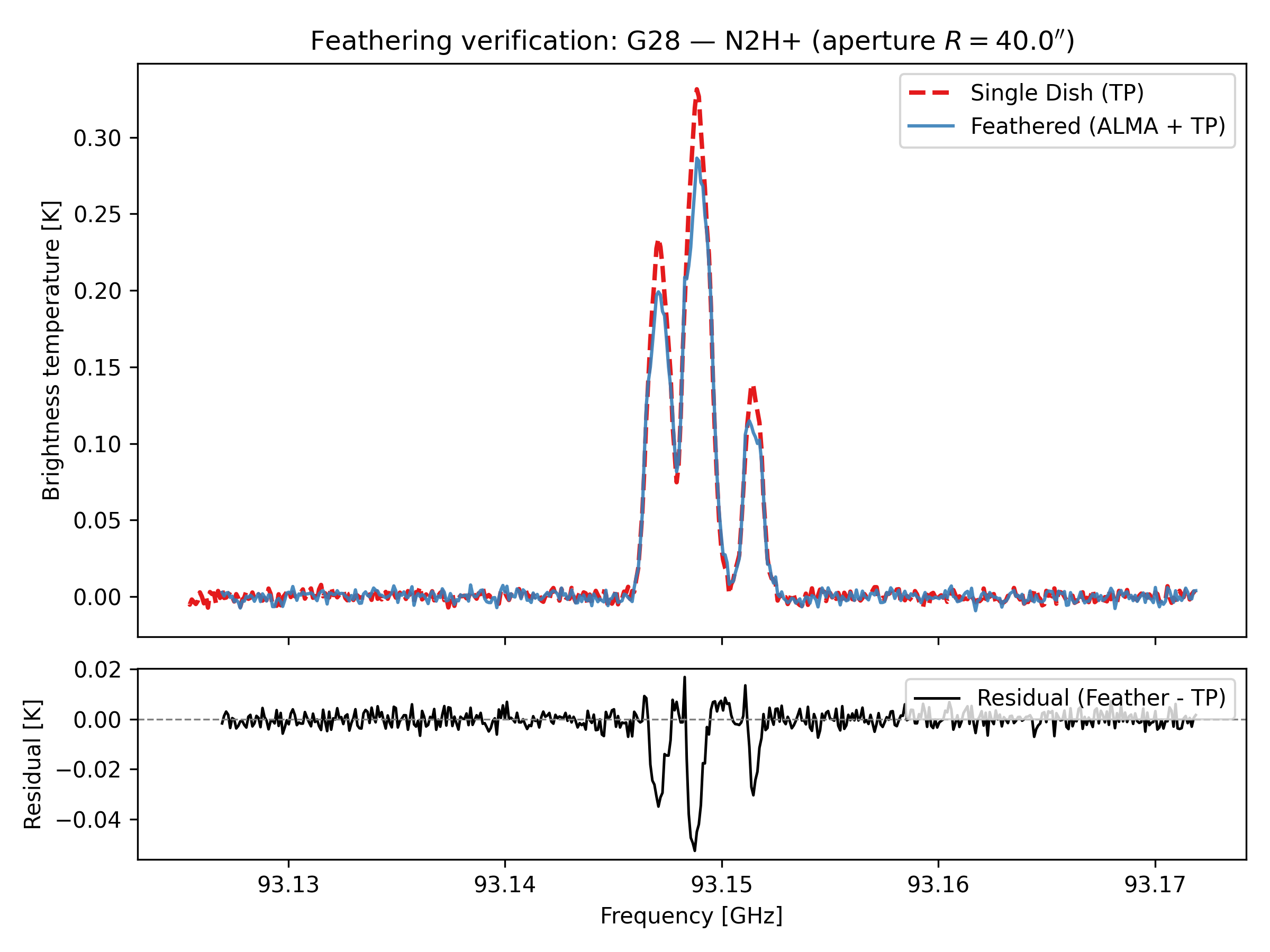}
    \includegraphics[width=0.41\linewidth]{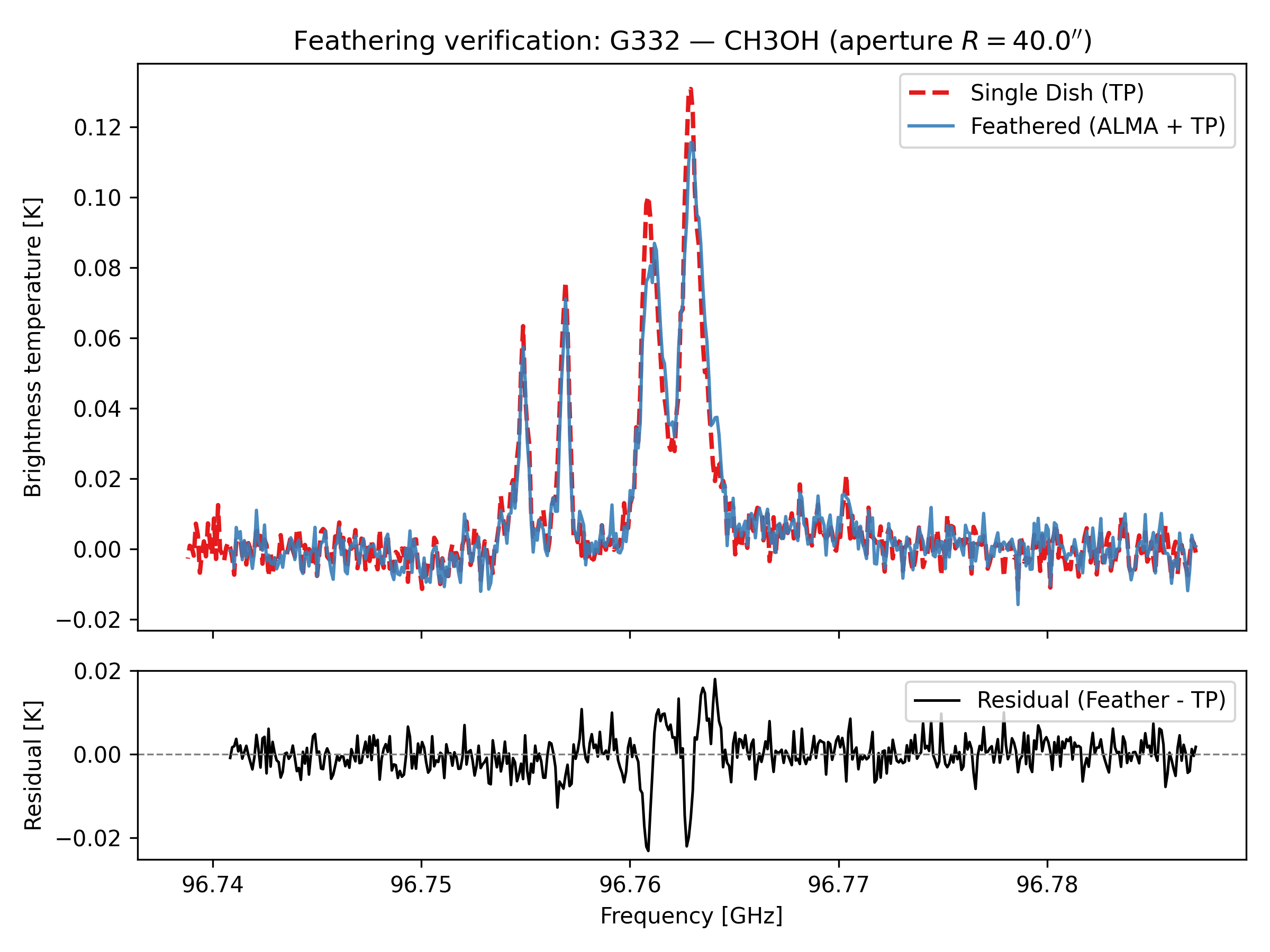}
    \hspace{0.1\linewidth}
    \includegraphics[width=0.41\linewidth]{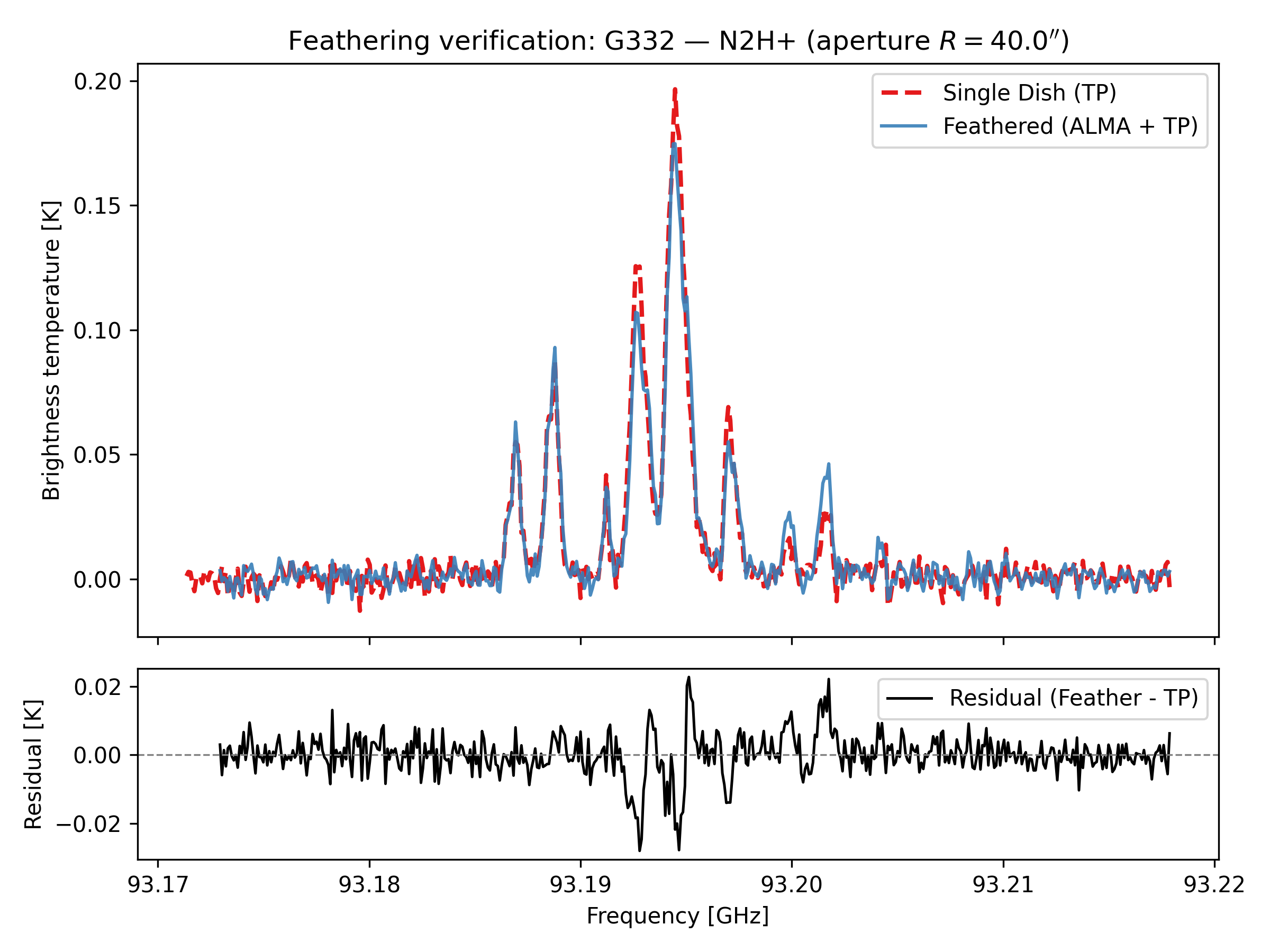}
    \caption{Comparison of the mean brightness temperature from the single-dish and the feathered cubes, computed over a $40\arcsec$ radius aperture to encompass the full single-dish beam. The bottom panels show the residuals (feathered - TP). The excellent agreement confirms that the data combination process recovers all extended flux without distorting the line profiles.}
    \label{fig:feather_check}
\end{figure*}

Figure~\ref{fig:feather_check} shows the spectral superpositions and residuals for $\mathrm{CH_3OH}(2_K-1_K)$ and $\mathrm{N_2H^+}(1-0)$ for all sources. The feathered spectra exhibit excellent agreement with the TP-only data. The negligible residuals (with maxima in the range 7.5-17.5\%) confirm that the feathering process successfully recovered the total flux measured by the single-dish telescope, preserving the true, unfiltered emission and kinematic structure of the clumps.

\FloatBarrier

    \section{Derived densities}\label{app:densities}

\begin{table}[!h]
    \centering
    \small
    \caption{Densities inferred for \gttt, in each of the regions where a spectrum has been extracted, for each velocity component.}
    \label{tab:G332.96_densities}
    \begin{tabular}{l l l r r r r l r}
        \toprule
        Region ID & RA & Dec & R & $V_{LSR}$ & FWHM & Best fit $n(\mathrm{H_2})$ & 67\% HPD & Rel. score \\
        & (H:M:S) & (D:M:S) & ($\arcsec$) & ($\mathrm{km\, s^{-1}}$) & ($\mathrm{km\, s^{-1}}$) & ($log_{10}(n(\mathrm{H_2}))$) & ($log_{10}(n(\mathrm{H_2}))$) & \\
        \midrule
        \gttt\_core\_cont\_01 & 16:18:32.53 & -50:24:59.26 & 2.0 & -67.96 & 1.06 & 5.0 & [4.7, 5.2] & 0.334 \\
        \gttt\_core\_cont\_02 & 16:18:32.32 & -50:24:54.86 & 2.0 & -66.53 & 0.62 & 4.7 & [4.3, 5.0] & 0.128 \\
        \gttt\_core\_cont\_03 & 16:18:32.01 & -50:24:59.46 & 2.0 & -66.50 & 0.59 & 4.6 & [4.2, 4.9] & 0.213 \\
        \gttt\_envelope\_01 & 16:18:33.05 & -50:25:02.26 & 3.0 & -66.39 & 0.59 & 5.0 & [4.7, 5.2] & 0.313 \\
        \gttt\_envelope\_02 & 16:18:30.90 & -50:24:42.26 & 5.0 & -65.97 & 0.88 & 4.6 & [4.3, 4.9] & 0.195 \\
        \gttt\_envelope\_03 & 16:18:30.02 & -50:24:31.86 & 5.0 & -66.34 & 0.70 & 4.7 & [4.3, 5.0] & 0.194 \\
        \gttt\_envelope\_04 & 16:18:33.14 & -50:24:50.46 & 5.0 & -66.89 & 1.14 & 4.7 & [4.3, 4.9] & 0.137 \\

        \bottomrule
    \end{tabular}
    \tablefoot{
        Core regions represent compact peaks (in continuum or CH$_3$OH), while envelope regions trace the ambient, diffuse gas of the parent clump.
        The columns list the Region ID, the centre coordinates and radius, the velocity of the component, the measured line FWHM, the best fit number density, its $67\%$ highest probability density (HPD) interval, and the reliability score, ranging from $0$ to $1$ and increasing for more reliable results.
    }
\end{table}

\begin{table}[!h]
    \centering
    \small
    \caption{Densities inferred for \gte, in each of the regions where a spectrum has been extracted, for each velocity component.}
    \label{tab:G28_densities}
    \begin{tabular}{l l l r r r r l r}
        \toprule
        Region ID & RA & Dec & R & $V_{LSR}$ & FWHM & Best fit $n(\mathrm{H_2})$ & 67\% HPD & Rel. score \\
        & (H:M:S) & (D:M:S) & ($\arcsec$) & ($\mathrm{km\, s^{-1}}$) & ($\mathrm{km\, s^{-1}}$) & ($log_{10}(n(\mathrm{H_2}))$) & ($log_{10}(n(\mathrm{H_2}))$) & \\
        \midrule
        \gte\_core\_ch3oh\_01 & 18:43:32.50 & -04:13:31.25 & 2.0 & 77.58 & 1.78 & 5.1 & [4.8, 5.3] & 0.261 \\
        \gte\_core\_cont\_01 & 18:43:31.44 & -04:13:20.05 & 2.0 & 79.34 & 0.93 & 4.9 & [4.6, 5.1] & 0.332 \\
        \gte\_core\_cont\_02 & 18:43:31.31 & -04:13:15.65 & 2.0 & 79.24 & 1.15 & 4.9 & [4.6, 5.1] & 0.394 \\
        \gte\_core\_cont\_03 & 18:43:31.11 & -04:13:20.25 & 2.0 & 79.44 & 1.37 & 4.8 & [4.5, 5.0] & 0.262 \\
        \gte\_core\_cont\_04 & 18:43:30.79 & -04:13:20.45 & 2.0 & 79.53 & 1.38 & 4.6 & [4.2, 4.8] & 0.250 \\
        \gte\_core\_cont\_04 & 18:43:30.79 & -04:13:20.45 & 2.0 & 81.32 & 0.44 & 4.7 & [4.3, 5.0] & 0.114 \\
        \gte\_core\_cont\_04 & 18:43:30.79 & -04:13:20.45 & 2.0 & 77.64 & 0.27 & 4.6 & [4.3, 5.0] & 0.084 \\
        \gte\_core\_cont\_05 & 18:43:31.19 & -04:13:18.25 & 2.0 & 79.21 & 1.23 & 4.7 & [4.5, 5.0] & 0.351 \\
        \gte\_core\_cont\_06 & 18:43:32.38 & -04:13:34.05 & 2.0 & 78.47 & 1.66 & 4.6 & [4.2, 4.9] & 0.173 \\
        \gte\_core\_cont\_06 & 18:43:32.38 & -04:13:34.05 & 2.0 & 81.96 & 0.63 & 4.7 & [4.3, 5.0] & 0.123 \\
        \gte\_core\_cont\_07 & 18:43:31.01 & -04:13:36.05 & 2.0 & 80.73 & 0.48 & 4.7 & [4.3, 5.0] & 0.180 \\
        \gte\_core\_cont\_07 & 18:43:31.01 & -04:13:36.05 & 2.0 & 78.09 & 1.96 & 4.7 & [4.2, 4.9] & 0.180 \\
        \gte\_core\_cont\_08 & 18:43:30.61 & -04:13:33.25 & 2.0 & 79.49 & 0.63 & 5.4 & [5.1, 5.7] & 0.208 \\
        \gte\_core\_cont\_08 & 18:43:30.61 & -04:13:33.25 & 2.0 & 81.02 & 0.90 & 4.7 & [4.3, 5.0] & 0.187 \\
        \gte\_core\_cont\_08 & 18:43:30.61 & -04:13:33.25 & 2.0 & 77.02 & 1.01 & 4.7 & [4.3, 5.0] & 0.107 \\
        \gte\_core\_cont\_08 & 18:43:30.61 & -04:13:33.25 & 2.0 & 68.71 & 0.88 & 4.7 & [4.3, 5.0] & 0.056 \\
        \gte\_envelope\_01 & 18:43:31.77 & -04:13:23.05 & 3.0 & 79.19 & 0.99 & 4.8 & [4.6, 5.0] & 0.396 \\
        \gte\_envelope\_02 & 18:43:30.40 & -04:13:03.05 & 5.0 & 79.21 & 1.22 & 4.8 & [4.6, 5.0] & 0.394 \\
        \gte\_envelope\_03 & 18:43:29.84 & -04:12:52.65 & 5.0 & 79.69 & 1.06 & 5.0 & [4.7, 5.3] & 0.212 \\
        \gte\_envelope\_04 & 18:43:31.83 & -04:13:11.25 & 5.0 & 78.57 & 0.97 & 4.9 & [4.5, 5.0] & 0.309 \\
        \gte\_envelope\_05 & 18:43:30.26 & -04:13:19.45 & 5.0 & 79.77 & 1.35 & 4.8 & [4.5, 5.0] & 0.319 \\
        \gte\_envelope\_06 & 18:43:31.83 & -04:13:32.45 & 5.0 & 80.19 & 1.56 & 4.6 & [4.4, 4.9] & 0.344 \\
        \gte\_envelope\_07 & 18:43:32.27 & -04:13:04.05 & 5.0 & 77.61 & 0.66 & 5.0 & [4.7, 5.2] & 0.295 \\
        \gte\_envelope\_07 & 18:43:32.27 & -04:13:04.05 & 5.0 & 80.34 & 0.55 & 4.7 & [4.3, 5.0] & 0.091 \\
        \gte\_envelope\_07 & 18:43:32.27 & -04:13:04.05 & 5.0 & 82.07 & 0.91 & 4.7 & [4.3, 5.0] & 0.075 \\

        \bottomrule
    \end{tabular}
    \tablefoot{
        Core regions represent compact peaks (in continuum or CH$_3$OH), while envelope regions trace the ambient, diffuse gas of the parent clump.
        The columns list the Region ID, the centre coordinates and radius, the velocity of the component, the measured line FWHM, the best fit number density, its $67\%$ highest probability density (HPD) interval, and the reliability score, ranging from $0$ to $1$ and increasing for more reliable results.
    }
\end{table}

\begin{table}[!h]
    \centering
    \small
    \caption{Densities inferred for \gft, in each of the regions where a spectrum has been extracted, for each velocity component.}
    \label{tab:G14_densities}
    \begin{tabular}{l l l r r r r l r}
        \toprule
        Region ID & RA & Dec & R & $V_{LSR}$ & FWHM & Best fit $n(\mathrm{H_2})$ & 67\% HPD & Rel. score \\
        & (H:M:S) & (D:M:S) & ($\arcsec$) & ($\mathrm{km\, s^{-1}}$) & ($\mathrm{km\, s^{-1}}$) & ($log_{10}(n(\mathrm{H_2}))$) & ($log_{10}(n(\mathrm{H_2}))$) & \\
        \midrule
        \gft\_core\_ch3oh\_01 & 18:17:23.53 & -16:25:12.93 & 2.0 & 36.00 & 2.03 & 5.3 & [5.2, 5.4] & 0.803 \\
        \gft\_core\_ch3oh\_01 & 18:17:23.53 & -16:25:12.93 & 2.0 & 40.41 & 0.82 & 5.6 & [5.4, 5.7] & 0.407 \\
        \gft\_core\_ch3oh\_02 & 18:17:21.43 & -16:25:08.13 & 1.6 & 40.97 & 1.67 & 5.2 & [5.0, 5.4] & 0.441 \\
        \gft\_core\_ch3oh\_03 & 18:17:23.03 & -16:25:02.53 & 2.0 & 39.75 & 1.62 & 5.3 & [5.2, 5.4] & 0.726 \\
        \gft\_core\_ch3oh\_03 & 18:17:23.03 & -16:25:02.53 & 2.0 & 34.63 & 1.82 & 5.1 & [4.9, 5.2] & 0.638 \\
        \gft\_core\_ch3oh\_03 & 18:17:23.03 & -16:25:02.53 & 2.0 & 36.76 & 0.95 & 5.4 & [5.2, 5.5] & 0.556 \\
        \gft\_core\_cont\_01 & 18:17:22.43 & -16:25:01.73 & 2.0 & 40.17 & 2.31 & 5.4 & [5.3, 5.6] & 0.614 \\
        \gft\_core\_cont\_02 & 18:17:22.29 & -16:24:57.33 & 2.0 & 41.42 & 0.99 & 5.6 & [5.5, 5.7] & 0.960 \\
        \gft\_core\_cont\_02 & 18:17:22.29 & -16:24:57.33 & 2.0 & 38.35 & 3.29 & 5.9 & [5.7, 6.0] & 0.705 \\
        \gft\_core\_cont\_03 & 18:17:22.08 & -16:25:01.93 & 2.0 & 41.09 & 0.80 & 5.3 & [5.1, 5.5] & 0.393 \\
        \gft\_core\_cont\_03 & 18:17:22.08 & -16:25:01.93 & 2.0 & 37.14 & 1.27 & 5.1 & [4.9, 5.4] & 0.259 \\
        \gft\_core\_cont\_04 & 18:17:21.75 & -16:25:02.13 & 2.0 & 40.40 & 1.21 & 5.2 & [5.0, 5.4] & 0.531 \\
        \gft\_core\_cont\_05 & 18:17:22.17 & -16:24:59.93 & 2.0 & 38.47 & 2.91 & 5.7 & [5.6, 5.9] & 0.685 \\
        \gft\_core\_cont\_05 & 18:17:22.17 & -16:24:59.93 & 2.0 & 40.25 & 0.83 & 4.6 & [4.3, 4.9] & 0.176 \\
        \gft\_core\_cont\_06 & 18:17:23.40 & -16:25:15.73 & 2.0 & 41.49 & 0.80 & 5.4 & [5.2, 5.6] & 0.380 \\
        \gft\_core\_cont\_07 & 18:17:21.99 & -16:25:17.73 & 2.0 & 40.74 & 1.08 & 5.6 & [5.5, 5.7] & 0.761 \\
        \gft\_core\_cont\_07 & 18:17:21.99 & -16:25:17.73 & 2.0 & 38.02 & 2.62 & 5.4 & [5.2, 5.6] & 0.345 \\
        \gft\_envelope\_01 & 18:17:22.78 & -16:25:04.73 & 3.0 & 40.22 & 1.00 & 4.8 & [4.6, 5.0] & 0.426 \\
        \gft\_envelope\_02 & 18:17:21.35 & -16:24:44.73 & 5.0 & 40.35 & 1.05 & 4.6 & [4.2, 4.8] & 0.238 \\
        \gft\_envelope\_03 & 18:17:20.76 & -16:24:34.33 & 5.0 & 40.36 & 1.23 & 4.6 & [4.2, 4.8] & 0.231 \\
        \gft\_envelope\_04 & 18:17:22.83 & -16:24:52.93 & 5.0 & 38.54 & 0.92 & 4.6 & [4.2, 4.8] & 0.220 \\

        \bottomrule
    \end{tabular}
    \tablefoot{
        Core regions represent compact peaks (in continuum or CH$_3$OH), while envelope regions trace the ambient, diffuse gas of the parent clump.
        The columns list the Region ID, the centre coordinates and radius, the velocity of the component, the measured line FWHM, the best fit number density, its $67\%$ highest probability density (HPD) interval, and the reliability score, ranging from $0$ to $1$ and increasing for more reliable results.
    }
\end{table}

\FloatBarrier

\section{Geometrical model for filamentary infall}\label{app:geometry}

\begin{figure}
        \centering
        \includegraphics[width=0.8\textwidth]{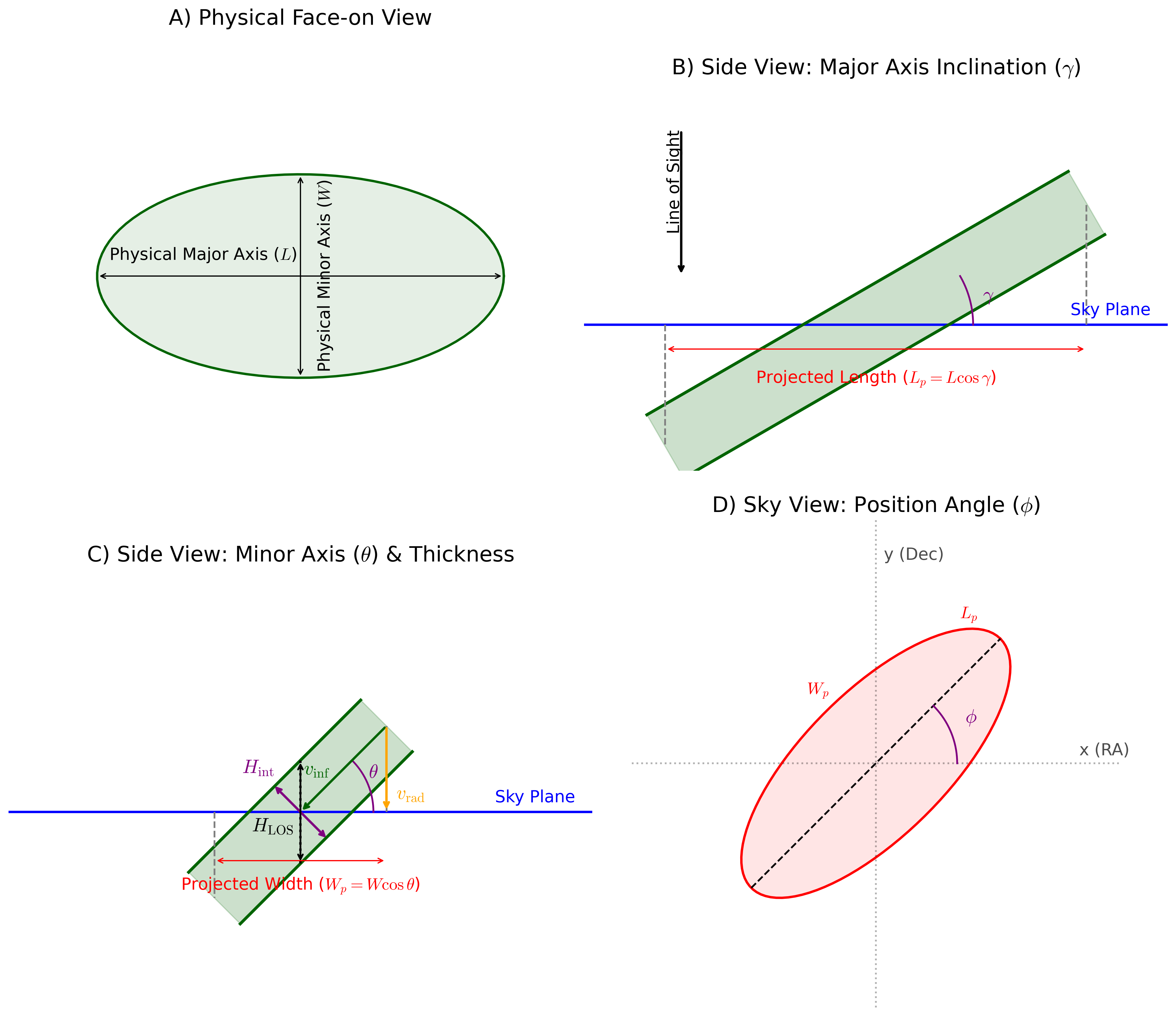}
        \caption{Schematic representation of the geometry used to estimate the filamentary infall rate.}\label{fig:geometry}
    \end{figure}

    As discussed in Sect.~\ref{sec:infall_method}, we approximate the filamentary clumps as inclined elliptical sheets. A simple picture of this simple model is provided in Fig.~\ref{fig:geometry}.

    We define the plane of the sky as the $xy$-plane. The elliptical sheet is inclined by an angle $\gamma$ along its major axis ($L$) and an angle $\theta$ along its minor axis ($W$). The physical (de-projected) dimensions are:
    \begin{equation}
        L = \frac{L_p}{\cos \gamma}, \quad W = \frac{W_p}{\cos \theta}
    \end{equation}
    where $L_p$ and $W_p$ are the observed axis diameters.

    \subsection{Infall velocity and timescale}
    The measured radial velocity gradient, $||\nabla\varv_{rad}||$, represents the change in the LOS velocity component over the projected radius ($W_p/2$). Therefore, the maximum LOS velocity difference from the edge to the centre is $\varv_{rad} = ||\nabla\varv_{rad}|| (W_p/2)$. The true physical infall velocity ($\varv_{inf}$) at the edge of the sheet is:
    \begin{equation}
        \varv_{inf} = \frac{\varv_{rad}}{\sin \theta} = \frac{||\nabla\varv_{rad}|| (W_p/2)}{\sin \theta} \label{eq:v_inf}
    \end{equation}
    The filamentary infall timescale is the time for material at the physical edge ($R = W/2 = W_p / (2 \cos \theta)$) to reach the centre:
    \begin{equation}
        \tau_{inf} = \frac{R}{\varv_{inf}} = \frac{W_p / (2 \cos \theta)}{||\nabla\varv_{rad}|| (W_p/2) / \sin \theta} = \frac{\tan \theta}{||\nabla\varv_{rad}||} \label{eq:t_inf_der}
    \end{equation}

    \subsection{Intrinsic thickness of the sheet}
    The parameter $H$ represents the intrinsic physical thickness of the sheet. We first estimate the path length of the material along the LOS ($H_{\mathrm{LOS}}$) by:
    \begin{equation}
        H_{\mathrm{LOS}} \approx \frac{N(\mathrm{H_2})}{n_{cl}(\mathrm{H_2})}\label{eq:los_depth}
    \end{equation}
    where $N(\mathrm{H_2})$ is the peak column density from ATLASGAL and $n_{cl}(\mathrm{H_2})$ is the mean volume density derived from our methanol fits (Table~\ref{tab:results_summary}). For our sources, this yields $H_{\mathrm{LOS}} \approx 0.15 - 0.2 \unit{pc}$.

    Because the sheet is inclined relative to the observer, this LOS path length is enlarged compared to the true thickness. The intrinsic thickness is recovered by applying the geometric correction:
    \begin{equation}
        H = H_{\mathrm{LOS}} \cos \gamma \cos \theta
    \end{equation}
    Using our assumed inclination angles ($\gamma = 30^\circ, \theta = 45^\circ$), this consistently yields an intrinsic thickness $H \approx 0.1 \unit{pc}$, which is used to calculate the physical volume of the sheet.

\end{appendix}

\end{document}